%% file: 00-Main.tex
\DocumentMetadata{}
\documentclass[sigconf]{acmart}
\usepackage[bottom]{footmisc}
\usepackage{colortbl}
\usepackage[font=small,skip=1pt]{subcaption}
\usepackage{enumitem}
\usepackage{tikz}
\usepackage{cleveref}
\usepackage{multirow}
\usepackage{xspace}
\usepackage{graphicx}
\usepackage{pdfpages}
\usepackage{xcolor}
\usepackage{balance}
\newcolumntype{?}{!{\vrule width 1pt}}
\newcolumntype{C}[1]{>{\centering\arraybackslash}m{#1}}
\newcolumntype{L}[1]{>{\raggedright\arraybackslash}m{#1}}

\definecolor{lightGrey}{RGB}{240,240,240}
\definecolor{darkGrey}{RGB}{200,200,200}
\definecolor{Grey}{RGB}{80,80,80}

\definecolor{revisedtext}{RGB}{0, 0, 0}
\newcommand{\revised}[1]{\textcolor{revisedtext}{#1}}

\newcommand{\emoji}[1]{%
    \raisebox{-0.3ex}{\includegraphics[height=0.95em]{emoji_images/#1.png}}\xspace
}

\AtBeginDocument{%
  \providecommand\BibTeX{{%
    \normalfont B\kern-0.5em{\scshape i\kern-0.25em b}\kern-0.8em\TeX}}}

\begin{document}

\author{Xi Zheng}
\affiliation{%
 \department{Department of Computer Science}
  \institution{City University of Hong Kong}
  \city{Hong Kong}
  \country{China}}
\email{zheng.xi@my.cityu.edu.hk}

\author{Zhuoyang Li}
\affiliation{%
 \department{Department of Computer Science}
  \institution{City University of Hong Kong}
  \city{Hong Kong}
  \country{China}}
\email{zhuoyanli4@cityu.edu.hk}

\author{Xinning Gui}
\affiliation{%
 \department{The College of Information Science and Technology}
  \institution{The Pennsylvania State University}
  \city{University Park, PA}
  \country{USA}}
\email{xinninggui@psu.edu}

\author{Yuhan Luo}
\affiliation{%
 \department{Department of Computer Science}
  \institution{City University of Hong Kong}
  \city{Hong Kong}
  \country{China}}
\email{yuhanluo@cityu.edu.hk}

\title{Customizing Emotional Support: How Do Individuals Construct and Interact With LLM-Powered Chatbots}

\renewcommand{\shorttitle}{Customizing Emotional Support}

\input{01-Abstract}
\input{Figures/01_interface}

\ccsdesc[500]{Human-centered computing~HCI design and evaluation methods}

\keywords{Emotional support, Chatbot, Wellbeing, Large language model, Prompt, Customization}

\maketitle
\input{02-Introduction}
\input{03-Related_Work}
\input{04-Study_Design_Goals.tex}

\input{05-ChatLab.tex}
\input{06-Method.tex}

\input{07-Findings-01.tex}

\input{08-Findings-02.tex}
\input{09-Discussion.tex}

\input{10-Conclusion.tex}

\begin{acks}
We thank our participants for their interest and contributions to this study and anonymous reviewers for their thoughtful suggestions. We also thank our colleagues Can Liu, Xiaoyu Zhang, Shengdong Zhao, Junnan Yu, and other members of the KLIC community for their feedback. The project was supported by City University of Hong Kong (\#9610597 and \#7005997).
\end{acks}

\bibliographystyle{ACM-Reference-Format}
\balance
\bibliography{Reference}

\end{document}

%% file: 01-Abstract.tex
\begin{abstract}
Personalized support is essential \revised{to fulfill individuals' emotional needs and sustain their mental well-being}. Large language models (LLMs), with great customization \revised{flexibility, hold promises to enable individuals to create their own emotional support agents}. \revised{In this work}, we developed ChatLab, \revised
{where} users \revised{could} construct LLM-powered chatbots with additional interaction features \revised{including} voices and avatars. \revised{Using a Research through Design approach, we conducted} a week-long field study followed by interviews and design activities \revised{(\textit{N} = 22), which} uncovered how participants created diverse chatbot personas for emotional reliance, confronting stressors, connecting to intellectual discourse, reflecting mirrored selves, etc. \revised{We found that} participants actively enriched the personas they constructed, shaping the dynamics between themselves and the chatbot to foster open and honest conversations. \revised{They also} suggested \revised{other} customizable features, such as integrating online activities and adjustable memory settings. Based on these findings, we discuss opportunities for enhancing personalized emotional support through emerging \revised{AI} technologies.

\end{abstract}

%% file: Figures/01_interface.tex
\begin{teaserfigure}
         \centering
         \includegraphics[width=1\textwidth]{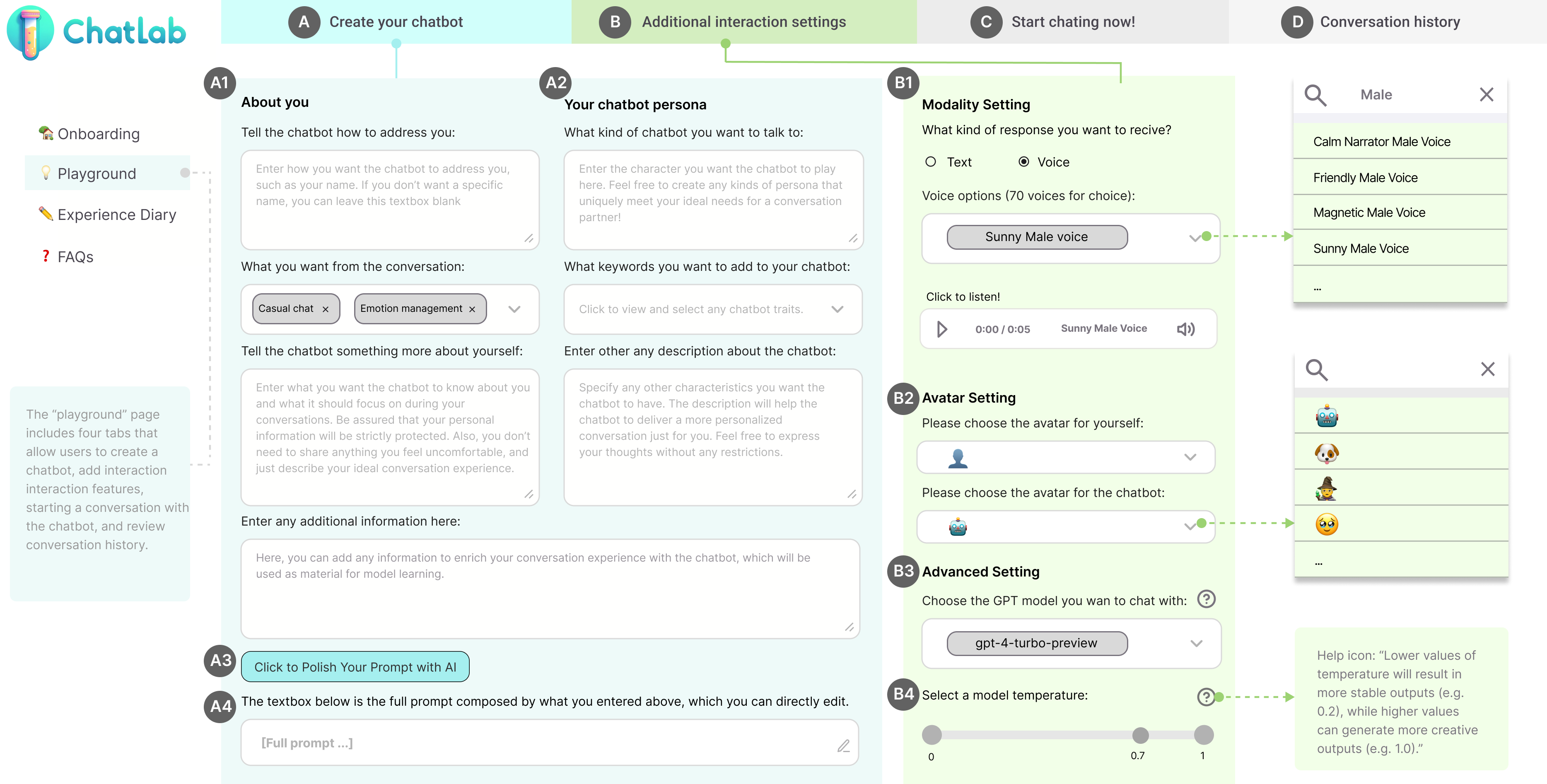}
          \caption{The four tabs in ChatLab's Customization and Conversation Playground: Chatbot customization (A), Additional interaction settings (B), Chatting (C), and Conversation history (D). In tab A, we designed multiple text boxes (optional) to encourage participants to provide more information about themselves (A1) and their desired chatbot persona (A2), along with an AI polish feature (A3) and the composed prompt that is editable (A4). In tab B, people can choose the chatbot's output modality (B1), avatar (B2), the LLM model applied (B3), and temperature (B4). The website is mobile responsive. The original interface was in Chinese, which we translated into English.}
          \label{fig:interface}
\end{teaserfigure}

%% file: 02-Introduction.tex
\section{Introduction}
Millions of individuals suffer from \revised{poor mental wellbeing due to lack of emotional support}~\cite{WHO2023Stress, ADAA2023Anxiety}. Some of them need a space to confide, some seek empathetic understanding, some want companionship, and others need practical guidance~\cite{lee2019caring, ding2023infrastructural, Gallup2023Lonely, sharma2020engagement}. These varying needs often warrant personalized support tailored to individuals' personal experience, sources of stress, etc~\cite{Cuijpers2016Personalized, simon2010personalized, Simm2016Personalized, schneider2015state}.
In the past few years, chatbots have emerged as an affordable tool to support those who struggle with mental well-being \revised{and even specific mental health conditions}~\cite{Lee2020Selfdisclosure, lee2019caring, hoiland2020hi, meng2021emotional}. While prior research has broadly explored different chatting styles to promote user engagement and \revised{self-disclosure}~\cite{Lee2020Selfdisclosure, lee2019caring, meng2021emotional}, little work has examined how chatbots can adapt to varying needs across individuals---or how individuals may customize a chatbot for their own emotional support.

The recent surge of large language models (LLMs) opened up opportunities for such chatbot design. In particular, the use of prompts---natural language instructions---to steer the model's output significantly lowered the barriers for lay people who are not AI experts to customize a chatbot persona by specifying its personality, communication style, and domain knowledge~\cite{pengfei_2023_prompt, Ha2024Clochat, characterai}. This flexibility allows people to take an active role in seeking support, potentially empowering them to articulate their emotional needs and receive personalized care through a cost-effective channel~\cite{song2024typingcure, li4875898human}.

Despite the potential, existing research largely focused on fine-tuning and prompting LLMs to emulate mental health professionals~\cite{wu2024sunnie, liu2023chatcounselor, chen2023llm} or exploring the use cases related to emotional support~\cite{song2024typingcure, li4875898human, jo2023understanding, Ha2024Clochat}. Yet, a notable gap remains in understanding how individuals leverage the ``customizability'' of LLMs to meet their unique needs by shaping model responses and the motivations driving their practices. Furthermore, prior work involving customizing LLM-powered chatbots primarily zeroed in on language outputs, offering limited insight into how customization of other chatbot attributes, such as voices and avatars, can augment user experiences and emotional well-being~\cite{Feine2019Taxonomy, deSouza2022Design}.

To address the abovementioned research gap, we took a ``Research through Design'' (RtD) approach, \revised{which uses design as an inquiry to investigate and understand complex user problems~\cite{zimmerman2007research, zimmerman2014research}. As part of RtD, researchers often have target users interact with existing artifacts or those they have created themselves, aiming to transform design solutions from ``the current state to a preferred state.'' In this process, the artifacts serve as a probe that elicits opportunities to advance the current design solutions from users~\cite{zimmerman2007research}. Following this approach, we designed and developed} ChatLab, a research prototype that allows people to construct \revised{and interact with} chatbots based on their conversation purposes (see Figure~\ref{fig:interface}). \revised{To empower individuals as active designers of their conversational partners, we carefully designed ChatLab's customization interface by including features such as open text hints and a diverse selection of avatar and voice options, providing users with rich options for customization.} We invited 22 participants experiencing moderate to high social loneliness to engage with ChatLab for seven to ten days, followed by interviews and design activities where participants \revised{shared their study experience and illustrated features that are not yet available to further enhance their interactions}.

\revised{Throughout} the study, participants constructed diverse chatbot personas to seek emotional reliance, confront stressors, connect to intellectual discourses, foster self-discovery and request therapeutic support. They also actively incorporated the voice and avatar options, along with other specifications of attributes, to enrich the constructed persona, shape relationship dynamics between themselves and the chatbot, and promote open and honest discussions. 
During the \revised{interviews and} design activities, participants came up with several ideas to enhance the chatbot's customizability beyond using prompts and pre-created attributes. They brought up mechanisms for AI to proactively learn about themselves through tracking their digital trace, physical environments, and emotional states. They also envisioned having the ability to manage past conversations and hoped for an AI assistant and community support to help lower their customization efforts.

\revised{Taking together participants' real-life experiences with their aspirations}, we discuss how chatbot customization enabled individuals to explore and reflect on their feelings, thoughts, and needs, as well as their desires to be emotionally connected to the chatbot. We also synthesized design implications to better support chatbot customization. 
This work contributes to the HCI community in two folds: (1) an empirical understanding of individuals' real-world practices in constructing and interacting with LLM-powered chatbots for daily emotional challenges; and (2) implications for designing effective emotional support tools tailored to individualized needs in the age of Generative AI.

\revised{It is noteworthy that, according to Keyes's dual continuum model~\cite{keyes2012promoting}, mental health problems and mental well-being are viewed as coexisting but independent. That is, poor mental well-being does not necessarily imply diagnosed mental health problems and vice versa.}
In this work, we focus on ``emotional support,'' which encompasses a broader range of \revised{mental well-being} assistance \revised{(e.g., stress coping, attentive listening)} and is less formal than mental health interventions~\cite{webMD2024}, as we do not intend to endorse LLMs to assist those with severe mental health issues. Thus, our findings should not be interpreted as clinical advice. 
\revised{However, these findings remain valuable for developing technologies to support those who experience poor mental well-being (e.g., social isolation).}

%% file: 03-Related_Work.tex
\section{Related Work}
In this section, we cover related work on existing ways tailoring emotional support to individualized needs and chatbot customization before and after the surge of Large Language Models (LLMs).

\subsection{Existing Approaches to Support Individuals' Emotional Well-Being} 

In clinical settings, supporting different individuals' mental health and well-being often warrants personalized care---identifying the key \revised{personal} characteristics (e.g., sociodemographic factors, sources of stress) that play parts in one's preferences and treatment outcomes to match the type of support~\cite{Cuijpers2016Personalized, simon2010personalized, schneider2015state}.
As an example, depending on a person's emotional reactivity (i.e., the intensity of responding emotionally to external stimuli), a health professional may choose between cognitive behavioral therapy (CBT) and other types of treatments, such as family-based therapy and mindfulness-based stress reduction~\cite{schneider2015state}.
\revised{This approach aligns with the concept of ``personalized medicine,'' which suggests that individual characteristics can guide treatment to maximize efficacy~\cite{schneider2015state}. Although in the context of mental well-being support, personal traits such as personalities and sociodemographics are less precise than disease-level factors in physical health, they can largely shape the effectiveness of care strategies and outcomes~\cite{schneider2015state, Cuijpers2016Personalized, simon2010personalized}.}

However, not everyone \revised{can access such tailored professional care} due to financial, time, and location constraints~\cite{brouwers2020social, WHO2019MentalDisorder, wainberg2017challenges}. In response, we have witnessed the emergence of numerous peer support networks and self-care applications, which enabled people to take an active role in managing their health and well-being~\cite{renwen_2021_customization, six2022effect, paul2021customizability, kim2019toward, gao2022taking}. For instance, Zhang et al. conducted a longitudinal study by deploying IntelliCare, a suite of mental health apps that individuals can selectively use, to people with depression or anxiety~\cite{renwen_2021_customization}. The study showed that offering a variety of intervention choices not only enhanced individuals' autonomy but also addressed their needs for encouragement, goal setting, and reflection across different situations~\cite{renwen_2021_customization}. 
Among \revised{these} existing support tools, chatbots have become increasingly prominent for their ability to engage individuals in natural language conversations, providing empathy and intimacy akin to human companions~\cite{Alaa2019Review, Aditya2019Review, Fitzpatrick2017CBT, Skjuve2021Companion, lee2019caring}.
More importantly, as an agent with humanlike attributes, chatbots can be customized through multiple aspects, such as communication style and visual appearance, which allows researchers to explore a broad range of customization opportunities~\cite{deSouza2022Design, Aditya2019Review, Alaa2019Review, Haque2023Overview, Xiao2007Customization}. 
For example, Xiao et al. enabled users to choose a chatbot's speed of speech and use of language (formal vs casual), which was shown to effectively improve their perceptions of the chatbot's performance~\cite{Xiao2007Customization}. Maia De Souza et al. analyzed the features of commercial chatbots designed to support depression management and highlighted that the customizability of the chatbot's avatar and personality are key features that influence the overall user experience~\cite{deSouza2022Design}.

While prior works revealed the potential of customizable chatbots to support individuals' emotional well-being, these works examined only limited and surface-level customization options (e.g., chatbot avatar, personality), which may not adequately address the intricate nature of individualized emotional needs. The limitations were largely due to technical challenges in providing flexible customization mechanisms before the surge of large language models (LLMs). As a result, our understanding of individuals' customization needs for emotional support, and the reasons behind their preferences still remains underexplored.

\vspace{-1.5mm}
\subsection{Chatbot Customization in The Age of Large Language Models (LLMs)}

In the past few years, the advent of Large Language Models (LLMs), such as OpenAI's GPT and Google's Gemini, has opened up new possibilities for building chatbots to support emotional well-being~\cite{chiang2024chatbot, jo2023understanding, Sharma2024Restructuring}.
Unlike in the pre-LLM era, where chatbots often operated within rigid and predefined conversation flows, these advanced models are equipped with unprecedented capabilities in natural language understanding, reasoning, and content generation~\cite{chiang2024chatbot, jo2023understanding, Sharma2024Restructuring}. 
Particularly, one of the excitements brought by LLMs is allowing users to employ~\textit{prompts}---natural language instructions---to steer a model's output, tailoring its tone, or even specifying its domain knowledge~\cite{pengfei_2023_prompt, Ha2024Clochat, zamfirescu2023herding, zamfirescu2023johnny, characterai}. This evolution enables people to customize LLM-powered chatbots with much more flexibility.
For instance, Ha et al. introduced Clochat, a platform that allows users to tailor LLM personas by selecting predefined options such as demographic details, domain knowledge, verbal cues, and visual cues~\cite{Ha2024Clochat}. They found that people enjoyed and engaged more with customized chatbot personas, regardless of the conversation topic~\cite{Ha2024Clochat}.
In a similar vein, platforms such as Character.AI have created various LLM-powered characters (e.g., psychologist, life coach, and fictional ``Elon Mask'') for users to interact with; it also enabled users to \revised{customize characters from scratch by specifying personality traits and unique backgrounds~\cite{characterai}. This capability empowers users to become active designers of their conversational partners rather than ``being personalized'' by models and developers}.

In a mental health context, Song et al. interviewed individuals about their use of LLM chatbots, such as ChatGPT and Pi~\cite{piai}, for support, and found that these chatbots fulfilled various roles, helping individuals with venting, comfort, routine conversations, and lifestyle advice~\cite{song2024typingcure}.
Based on their findings, the researchers discussed how AI mental health support tools aligned or failed to address people's emotional distress, calling for balancing user agency over the interactions and their therapeutic growth~\cite{song2024typingcure}.
Similarly, Li et al. conducted a social media analysis on how people sought mental health support from ChatGPT and discovered a variety of prompts that they employed, such as strategically guiding GPT to act more emotionally intense despite OpenAI's restrictions~\cite{li4875898human}.
While these works highlight how the adaptability of LLMs empowered people to actively seek mental health support through natural language interactions, they did not look into how people navigate these interactions and the motivations behind their practices. Additionally, existing work largely focused on customization practice on language outputs, with little understanding of how people may customize other chatbot attributes such as voice and avatars to enrich the interaction \revised{experiences}~\cite{Feine2019Taxonomy, deSouza2022Design}.

In this light, we set out to examine how individuals customize an LLM-powered chatbot for emotional support, aiming to delve into individuals' customization practice, along with the interplay between their practices, emotional status, interaction experiences, and expectations for LLMs in everyday lives.

%% file: 04-Study_Design_Goals.tex
\section{ChatLab: A Prototype for Research Exploration}
Moving beyond traditional formative study approaches, we followed the ``Research through Design'' (RtD) approach~\cite{zimmerman2007research, zimmerman2014research} to design and build ChatLab, a research prototype that enables people to create customized chatbots for emotional support. \revised{Here, ChatLab serves as a medium to solicit experiences and gather design requirements for customizing chatbots to address emotional needs. }
By engaging people with ChatLab, we aimed to understand not only the specific options they customize but also how their emotional states influence their customization practice in real-world settings. 
Specifically, we ask two research questions (RQs): 

\begin{itemize}
    \item \textbf{RQ1}. How do individuals construct and interact with a customizable chatbot powered by large language models (LLMs) for emotional support? 
    \item \textbf{RQ2}. What design opportunities \revised{could potentially enhance the customizability of LLM-powered chatbots in meeting} individualized emotional needs? 

\end{itemize}

In this section, we describe the design rationales of ChatLab and its interface components.  
Note that we did not intend to evaluate ChatLab for its effectiveness in supporting individuals' emotional needs; instead, we aim to leverage it \revised{as part of the RtD approach} to answer the above research questions. 

\subsection{Design Rationales}

\subsubsection{DR1: Facilitating chatbot customization while maintaining user autonomy}
As mentioned in prior work~\cite{zamfirescu2023johnny, zamfirescu2023herding, li4875898human}, although the flexibility of prompts lowers the barriers to customizing LLM responses, it remains challenging for lay people who are not AI experts to compose effective prompts. To mitigate the learning curve and customization efforts, existing platforms, such as Character.AI and Poe, have offered rich lists of pre-created personas~\cite{characterai, poe} and customization options (e.g., demographics, personality keywords)~\cite{Ha2024Clochat}. However, people may be inclined to select from the available options rather than exploring their own ideas.
To facilitate LLM customization reflecting individuals' desired outputs without constraints, we need to carefully balance the amount of guidance to provide and user autonomy. This involves offering supportive resources, such as hints, examples, and prompt editing support, while encouraging people to express their unique needs and preferences.

\subsubsection{DR2: Expanding customization possibilities to enrich user experience} 
While text-level customization allows people to tailor the chatbot's use of language and communication style to their preferences, these verbal cues are not the only factors shaping individuals' conversational experience.
Non-verbal cues such as visual appearance and voice interaction are also important and can further complement the personas people create and enrich their conversation engagement~\cite{Feine2019Taxonomy, Xiao2007Customization}.
Therefore, we aim to expand the customization options beyond the text level by incorporating the chatbot's avatar and voice tones into ChatLab.

\subsubsection{DR3: Encouraging active customization for research exploration} 
As shown from prior research, individuals' emotional needs vary across time and occasions~\cite{mortenson1999cultural}. Thus, their desired conversation partner or its communication style may change as they experience different feelings and events. However, people may not actively customize new personas without appropriate \revised{guidance}, due to insufficient inspiration or interaction  efforts~\cite{renwen_2021_customization}.
To collect diverse customization samples in addressing our RQs, we thus need to encourage active customization throughout the study period.

%% file: 05-ChatLab.tex
\subsection{ChatLab Design Components}
To address these design rationales, we designed ChatLab, a mobile-responsive website that allows people to customize and interact with LLM-powered chatbots. The website has four key components: an onboarding page, customization and conversation playground, experience diary, and FAQs, which we elaborated on below.

\subsubsection{Onboarding Page and FAQs}

To streamline the onboarding process, we developed the Onboarding page and the FAQs page, respectively, to familiarize people with customization options powered by LLMs. 
After login, people are first directed to the Onboarding page, where they are greeted with an overview of the study objectives, complemented by a few dialogue examples with chatbots in different modalities (DR1). This page also features some avatars and voice samples to showcase additional interaction settings.

We designed the FAQs page to address common questions that non-AI experts might have (based on our pilot studies mentioned in Section~\ref{pilot}), focusing on two aspects. The first covered questions related to LLMs, such as what large language models are, what a ``prompt'' entails, and tips for creating effective prompts (DR1). 
For instance, to clarify the concept of prompt, we provided the following description: ``\textit{A prompt is like a food order you described to a waiter at a restaurant. Just as you detail your preferences to receive the dish you desire, you can guide the model to generate the content you need by sending it a clear prompt}.'' 
The second is about our study logistics and the research team's contact information.

\subsubsection{Customization and Conversation Playground}

When people enter the Customization and Conversation Playground, they are presented with four tabs as shown in Figure~\ref{fig:interface}: 
chatbot customization (A), additional interaction settings (B), chatting (C), and conversation history (D).

To encourage people to provide as much information as possible about their conversation expectations (DR2), we structured the customization flow using a `template' consisting of multiple text boxes, each highlighting specific information about their backgrounds (A1) and chatbot persona (A2).
Each text box was accompanied by hints designed to be inspiring and open-ended, assisting those who were new to LLMs. All the text entered will collectively compose the full prompt (A4) sent to the language model in the form of a bullet list, which people can also edit. Additionally, we designed an AI polish feature (A3), which was prompted to be ``\textit{an LLM expert who specialized in translating user-created content into a coherent prompt that covers every aspect of user request and is easier for GPT models to understand}'' to help people better organize and phrase their prompts with GPT-4. 
\revised{Note that since our goal is to let people articulate their own emotional needs, we intentionally avoided specific instructions and kept the customization space open to not lead individuals' customization behaviors or limit their creativity (DR1).}

In tab B, we provided additional interaction settings that allow people to customize the chatbot beyond what a prompt can offer (DR2).
These included the chatbot's output modality (B1) and avatar (B2), which have been shown as important customization options in prior research~\cite{Feine2019Taxonomy}. 
To align with the communication habits of mainland Chinese participants, our voice library included 70 distinctive Mandarin voices (DR2): six Chinese characters in OpenAI~\cite{openai} and 64 other characters from TikTok~\cite{tiktok}. The latter provided diverse voice options, including different genders (e.g., gentleman, considerate sister), ages (e.g., little kid, elderly), and other cartoon or movie characters (e.g., Crayon Shin-chan)~\cite{tiktok}.
As for avatar options, we turned to the widely used emoji set (77 options) because they can represent various moods, professions, animals, plants, and symbols, and most people are familiar with possible options. By default, the chatbot avatar is~\emoji{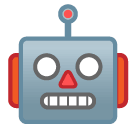} (a robot) and the user's avatar is~\emoji{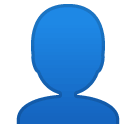} (a user silhouette).  We used emoji images from online resources~\footnote{Emoji images obtained from \url{https://emoji.aranja.com/} and \url{https://uxwing.com/}.} to generate emojis presented in this paper. Note that the actual appearance of emojis may vary depending on the user's device and browser.
We did not support voice input, mainly because it is often prone to recognition errors, which can greatly affect user experience~\cite{luo2022notewordy}.
Besides, people can select various GPT models (B3) to be applied in the conversation (i.e., GPT-4 Turbo, GPT-4, and GPT-3 models), along with the temperature setting (B4) that controls the randomness and creativity of the chatbot's responses.

Once all settings are configured, people can proceed to interact with the chatbot (C). 
During the conversation, people could always adjust the customized components in tabs A and B and then continue with the conversation. They could also switch between voice and text output as they like. Additionally, they can revisit their previous customization settings and review their conversation history in the Conversation History window (D).

\subsubsection{Experience Diary}
To better capture people's situational experiences during their customization and conversations with the chatbot, we incorporated an Experience Diary page where they were prompted to answer several questions about their daily experiences (DR3). This design was inspired by previous studies that utilized a diary to gather qualitative insights into user behavior and emotional responses over time~\cite{luo2020tandemtrack, luo2021foodscrap, blair2018onenote}.
The diary questions covered four aspects \revised{(see Figure~\ref{fig:conversations} for details)}: (1) a description of the conversation scenario or what motivated the person to interact with the chatbot; (2) the parts of the conversation that make them feel helpful or memorable, if any; (3) the customized setting of the chatbot and reasons of the settings; and (4) any additional thoughts.

\input{Figures/04_conversation_examples}

\subsection{Implementation}

ChatLab is built upon Streamlit, a Python framework for delivering interactive apps that embed AI models~\cite{streamlit}. The chatbot was powered by GPT models developed by OpenAI~\cite{openai}, following the guidelines of LangChain to insert the user-created prompts into the language models~\cite{langchain}.
We utilized the text-to-speech (TTS) APIs from OpenAI~\cite{openai} and Volcano Engine~\cite{tiktok} for the chatbot's voice output.
This architecture enables flexible integration of multiple APIs, while allowing people to seamlessly customize and interact with a customizable chatbot.
All the interaction data were securely stored at Firebase~\cite{firebase}. Access to the data is exclusively granted to the research team.

%% file: Figures/04_conversation_examples.tex
\begin{figure*}

         \centering
         \includegraphics[width=1\textwidth]{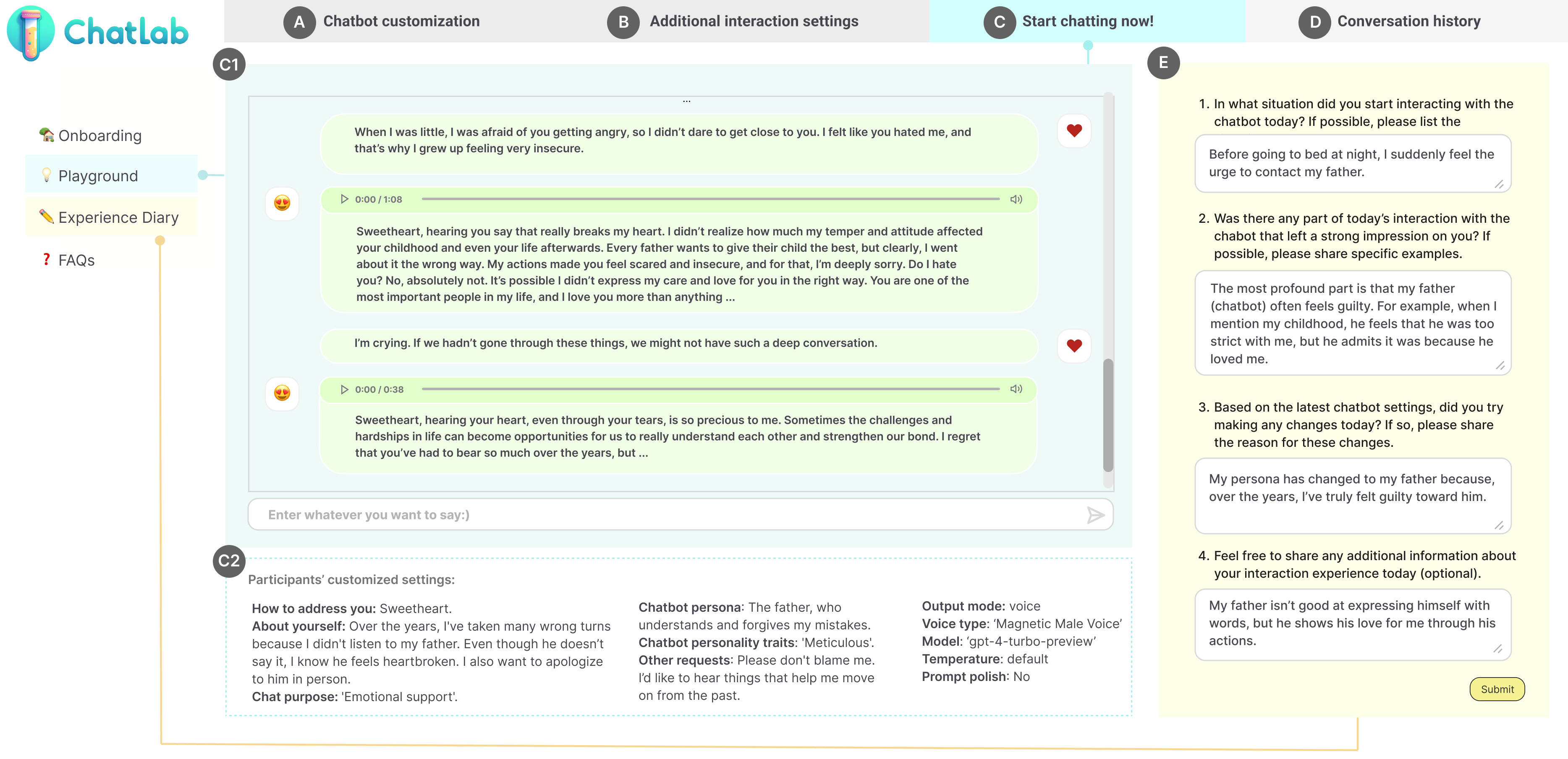}
          \caption{\revised{The Chatting interface (C1) with an example of participant P22's customized settings (C2) and one of their Experience Diary entries (E). Note that the participant chose voice as the interaction mode; for convenience, we transcribed the audio into text. The original interface, conversations, and diary entries were in Chinese and were later translated into English.}}
          \label{fig:conversations}

\end{figure*}

%% file: 06-Method.tex
\section{Method}
We first conducted a pilot study to examine the feasibility of deploying ChatLab in real-life situations and then proceeded to a formal study consisting of three phases: an online study tutorial, a field study involving interaction with ChatLab, and a post-study interview along with a design activity. The study was approved by the university's ethics review committee.

\subsection{Pilot Study}~\label{pilot}
We invited six participants who are interested in using AI to manage their emotional well-being to interact with ChatLab on a daily basis for a week.
The participants were all university students (three undergraduate and three graduate), and three of them had prior experiences with LLMs.
Next, they joined a two-hour focus group to share their experience with ChatLab. Their participation was voluntary without compensation.

Overall, participants created distinctive personas on ChatLab (e.g., a famous King in history or a virtual lover) and found the conversations engaging. However, participants pointed out several usability issues, which we later improved: 1) adding an extra input field on the chatbot customization tab to allow users to enter additional descriptions; 2) making the full prompt editable so that people do not need to follow the template we provided or revise the composed prompts; 3) condense some hints provided in the text boxes and avoid leading people towards particular options; 4) adding a conversation history tab for people to revisit their previous prompts and dialogues, which can be used as sources for future customization; 5) adding a frequently asked questions (FAQs) page about LLMs, prompt engineering, and tips to make effective prompts.
\revised{Besides, by examining the conversation logs of pilot participants, we did not identify any inappropriate or harmful output from their customized chatbots}.

\input{Tables/participants}

\subsection{Formal Study}
\subsubsection{Participants \& recruitment}~\label{participants}
We distributed the study call via several popular social media platforms in Mainland China, including Red, Hupu, Douban, and Weibo. To meet our inclusion criteria, individuals must be (1) above 18 years old; (2) facing some emotional struggles at the moment; and (3) at a moderate or high level of social loneliness.
In the screening questionnaire, we employed the UCLA Loneliness Scale~\cite{UCLA@Russell1978} to assess the levels of individuals' social loneliness, which has been widely used in prior research~\cite{KILLGORE2020loneliness, Aydin2021loneliness, Jamil2023loneliness}.
As mentioned earlier, we did not specifically look for participants with diagnosed mental health disorders, as we focused on broad emotional support.
Individuals experiencing social loneliness---who may not have formal diagnoses---are in need of various emotional support~\cite{wang2018associations} and often show greater openness to AI companions~\cite{crawford2024artificial}.
\revised{For safety considerations, we excluded participants who reported having severe mental health conditions (e.g., currently on medication, having a formal diagnosis, or undergoing professional treatment)}.

Among the 149 individuals who responded to our screening questionnaire, 30 met the above inclusion criteria, 27 attended our tutorial session, 26 completed the field study, and 22 attended the post-study interview. In the end, we included these 22 participants in our data analysis---they were 12 females and 10 males with an age range from 21 to 40 (\textit{M} = 25). Among them, 14 reported having limited or no experience with LLMs, and others indicated some familiarity with popular LLMs, such as ChatGPT or Bard. 
All of them reported a social loneliness score above 40 out of 80, indicating a moderately high level of social loneliness. Table~\ref{tab:ParticipantInfo} provides an overview of their backgrounds.
Each of these participants received RMB 300 (approximately USD 42) as compensation for their participation. 
The other five participants who attended the tutorial session, while not included in our data analysis, received RMB 100 (approximately USD 14) as a thank for their time.

\subsubsection{Pre-study tutorial}
Before the study, participants were invited to join an online tutorial via \revised{Tencent Meeting}. During the tutorial, we introduced the study procedure and provided instructions on how to interact with ChatLab. We explicitly explained how the text entered in each textbox on the customization interface is organized to form the ``prompt'' sent to the LLM and demonstrated various voices from different virtual characters. Next, we shared a set of distinctive prompting strategies collected from the pilot study. We clarified that participants were not required to create different personas during the study, but we encouraged them to be creative and explore various ways to construct a conversation partner that can best support their emotional well-being. \revised{We also emphasized the limitations of LLMs, advising participants not to seek medical diagnoses or clinical advice from the chatbots they customized.}
The tutorial lasted about 30 minutes and was organized in a group session, allowing multiple (one to five) participants to join together. 

\subsubsection{Field study}
After the tutorial, we sent each participant a unique username and password to access ChatLab. The next day, they started using ChatLab on a daily basis for seven to ten days, depending on their availability for the post-study interviews. 
Each day, participants were supposed to customize one or multiple chatbots on ChatLab based on their mood and emotional needs at specific times and then interact with the chatbots to share, express, or ask for anything they wish. They could also continuously adjust their customized settings during the conversations.
In the meantime, participants were asked to submit at least one diary entry to share their experience of customizing and interacting with the chatbot. To keep them engaged, our research team also sent each participant daily reminders through their preferred contact method.
As a minimum requirement, participants are expected to submit five diary entries during the study to proceed to the next phase.
\revised{This one-week field study served as a probe to help participants familiarize themselves with the capability of LLMs and prepare them for the upcoming interviews and design activities}.

\subsubsection{Post-study interview \& design activity}
In the last phase of the study, all participants who completed the field study with ChatLab took part in the post-study interview, except for two who declined the invite and two ``fraudulent'' participants who were later found to have provided deceptive information in the screening questionnaire to meet the inclusion criteria~\cite{panicker2024understanding, roehl2022imposter}. As mentioned earlier (Section~\ref{participants}), their data were therefore excluded from analysis.

The activities were online and took about 90 minutes. First, participants discussed their experiences using ChatLab. With participants' consent, we logged into ChatLab with their usernames and shared our screen so that they could point to specific prompts and conversations for discussion.

Next, \revised{each participant engaged in a design activity to reflect on their experiences of using Chatlab and envision their ideal conversation partner for emotional support. This activity is intended for participants to go beyond their current interactions with ChatLab and articulate their ideal experience to custmoize and interact with an intelligent chatbot.}
To facilitate the design activity online, we used Excalidraw~\footnote{\url{https://excalidraw.com/}}, a website that allows multiple people to draw and take notes on a virtual whiteboard at the same time (see Figure~\ref{fig:design-examples} for examples drawn by participants).
The design activity followed a ``think-aloud'' format, where participants articulated their thoughts while drawing on the whiteboard. The research team acted as a moderator to clarify their design ideas and assist them in adding notes to the designs when necessary. The original interviews were conducted in Chinese. We audio-recorded the entire sessions, which were later transcribed into text and then translated into English.

\subsection{Data Analysis}
Our data analysis centered on participants' customized chatbots, their interview transcripts, and designed artifacts (with notes), complemented by their conversation logs and diary entries. 
\revised{For each participant, we extracted their customization settings, including the descriptions about themselves, the chatbot persona, choice of avatars and voices, and statistics regarding their conversation rounds and time spent on these conversations.}

\input{Tables/behavior}

\revised{For interview transcripts and designed artifacts}, we took a bottom-up approach \revised{and} followed the stages of thematic analysis~\cite{braun2006using}. First, three researchers independently read through the interview transcripts and design artifacts of the same four participants to familiarize themselves with the data, and extracted excerpts deemed to be relevant and interesting. 
For each data excerpt, we labeled them with analytically meaningful codes such as ``\textit{multifaceted emotional needs including unconditional support, sincere advice, tolerance, and companionship}'' and ``\textit{prefer local accents that are similar to their own background}.'' During this process, we referred to the corresponding conversation logs and diary entries to ensure our interpretations were relevant to the contexts of the participants' interactions.
Then, the three researchers discussed their initial codes and resolved discrepancies through regular meetings. 
In parallel, two researchers divided up the work to code the remaining data following the same procedure. During this process, We continuously compared the newly emerged code with the previous codes, which eventually led to a total of 693 codes. 
Next, the entire research team worked together to develop higher-level themes based on the initial codes. 
We grouped similar codes together, aiming to solidify the underlying motives and rationales associated with the chatbot customization and emotional needs.

In this study, we focused on qualitative insights instead of quantitative measures as we sought an in-depth, contextual, and nuanced understanding of how and why individual participants customize chatbots for emotional support in particular ways, which can be difficult to quantify with numbers. To protect participants' privacy, we anonymized all the data and used P\# to denote them.

%% file: Tables/participants.tex

\begin{table*}[htbp]
 \sffamily\smaller
 \def\arraystretch{1}
 \centering
 \caption{Information about participants' backgrounds, their prior experience with LLMs, and the main emotional struggles they shared during the interview.
 }
 \label{tab:ParticipantInfo}
 \resizebox{1\textwidth}{!}{
 \begin{tabular}{!{\color{black}\vrule}  L{0.025\textwidth} | L{0.065\textwidth} |  L{0.325\textwidth} | L{0.14\textwidth} | L{0.14\textwidth} | L{0.25\textwidth}!{\color{black}\vrule}} 
 \hline
 \rowcolor{lightGrey}
 \small \textbf{ID} & \small \textbf{Age/ Gender} & \small \textbf{Occupation} & \small \textbf{Social loneliness} &\small \textbf{Prior experience with LLMs} & \small \textbf{Main struggle(s)} \\ 
 \hline
 P1 & 21/F & Management undergrad student & High & Limited & Social isolation, desire for companionship, eating disorder\\ \hline
 P2 & 28/F & Philosophy grad student & Moderate & Limited & Academic challenges \\ \hline
 P3 & 29/F & Philosophy grad student & Moderately High & Limited & Self-blaming, entanglement in a romantic relationship \\ \hline
 P4 & 25/F & Finance grad student & Moderately High & Some & Academic challenges\\ \hline
 P5 & 29/M & Product manager (IT) & Moderately High & Some & Diagnosed depression, work stress, existential crisis\\ \hline
 P6 & 26/M & Sales (real estate) & High & None & Social isolation, lingering attachment to the past relationship\\ \hline
 P7 & 23/M & Trade coordinator (international trade) & High & Some & Social anxiety\\ \hline
 P8 & 28/M & Engineer (alternative energy) & High & Limited & Social isolation\\ \hline
 P9 & 21/M & Automation student (junior college) & High & Limited & Social isolation, desire for companionship\\ \hline
 P10 & 22/ M & Mechanical engineering undergrad student & High & None & Emotional distress from the past relationship\\ \hline
 P11 & 23/F & Civil engineering grad student & Moderately High & Some & Low self-esteem\\ \hline
 P12 & 27/F & Materials science grad student & Moderately High & Limited & Social isolation, desire for companionship \\ \hline
 P13 & 23/F & Statistics grad student & Moderate & Some & Anxious about future employment\\ \hline
 P14 & 25/F & Operation specialist (e-commerce)& Moderately High & Limited & Desire for companionship\\ \hline
 P15 & 40/M & Frontline worker (state-owned enterprise) & High & None & Social isolation, desire for companionship\\ \hline
 P16 & 23/M & EE undergrad student & High & Limited & Social isolation, desire for companionship\\ \hline
 P17 & 21/M & EE student (junior college) & High & Some & Social isolation, desire for companionship\\ \hline
 P18 & 20/F & Nursing undergrad student & Moderately High & None & Family conflicts \\ \hline
 P19 & 21/F & Geography undergrad student & High & Limited & Academic challenges, family conflicts\\ \hline
 P20 & 21/M & Environmental science undergrad student & Moderately High & Some & Desire for companionship\\ \hline
 P21 & 23/F & Accounting undergrad student & High & Limited & Academic challenges\\ \hline
 P22 & 34/F & Secondary school teacher & High & Limited & Family conflicts, financial burden, depression\\ \hline
 \end{tabular}}
\end{table*}

%% file: Tables/behavior.tex

\begin{table*}[t]
 \sffamily
 \def\arraystretch{1.1}
 \centering
 \caption{\revised{Participants' customization settings during the study, including total conversation rounds, time spent per session in minutes (with mean and standard deviation), examples of chatbot personas created by participants with corresponding conversation rounds, and percentages of rounds with customized avatars and voice outputs with corresponding examples.
 }}
 \label{tab:Behaviour}
\resizebox{1\textwidth}{!}{
 \begin{tabular}
 {!{\color{black}\vrule} L{0.025\textwidth} | L{0.05\textwidth} L{0.14\textwidth}L{0.32\textwidth}  L{0.28\textwidth}  L{0.37\textwidth} !{\color{black}\vrule}} 
 
 \hline
 \rowcolor{lightGrey}
 \small \textbf{ID} & \small \textbf{Total rounds}  &\small \textbf{Time spent per session} & \small \textbf{Chatbot persona examples (rounds)} & \small \textbf{Rounds with customized avatars (chatbot avatar : participant avatar)} & \small \textbf{Rounds with customized voice}\\ 
 \hline
 P1 & 59 & M = 18, SD = 15 & An electronic dog (59) & 100\% (e.g., \emoji{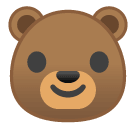} : \emoji{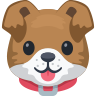}) & 66\% (e.g., lazy goat$^a$) \\ \hline
 P2 & 111 & M = 26, SD = 38 & A friend (55), a companion chatbot (28)  & 87\% (e.g., \emoji{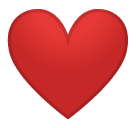} : \emoji{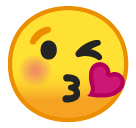}, \emoji{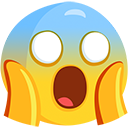} : \emoji{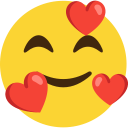}) & 0\% \\ \hline
 P3 & 59 & M = 15, SD = 11  & Their crush (21), an AI version of themselves (19), Musician Schubert (12) & 92\% (e.g., \emoji{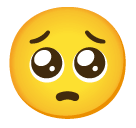} : \emoji{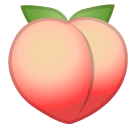}, \emoji{pleading-face} : \emoji{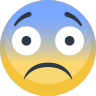}) & 0\% \\ \hline
 P4 & 148 & M = 37, SD = 41  &  A friend (49), their PhD supervisor (24), panda Huahua$^b$ (22), a cute daughter (12) & 99\% (e.g., \emoji{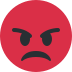} : \emoji{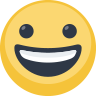}, \emoji{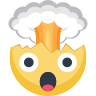} : \emoji{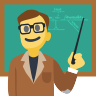}) & 18\% (e.g., lazy goat, endearing child) \\ \hline
 P5 & 121 & M = 14, SD = 13 & A psychologist (31), Sartre$^d$ (29), a patient who is diagnosed with anxiety (22), Heidegger$^d$ (18) & 0\% & 7\% (e.g., calm and mature male voice) \\ \hline
 P6 & 62 & M = 6, SD = 1 & No customization (28), a scholar (16), their ex-partner (11), a virtual girlfriend (7) & 18\% (e.g., \emoji{pouting-face} : \emoji{robot}) & 39\% (e.g., lazy goat, Alloy$^c$, Chongqing boy's voice$^e$) \\ \hline
 P7 & 46 & M = 4, SD = 1 & No customization (38), a lover (8) & 17\% (e.g., \emoji{grinning-face} : \emoji{grinning-face}) & 7\% (e.g., Alloy) \\ \hline
 P8 & 44 & M = 6, SD = 4 & A friend (28), no customization (9), a lover (7) & 0\% & 48\% (e.g., gentle lady, intellectual female voice) \\ \hline
 P9 & 50 & M = 13, SD = 7 & A friend (24), no customization (15), a family member (11) & 0\% & 0\% \\ \hline
 P10 & 31 & M = 6, SD = 1  & An AI psychologist (8), god (6), a wellness coach specialized in emotion regulation (6) & 100\% (e.g., \emoji{grinning-face} : \emoji{robot}) & 19\% (e.g., lazy goat) \\ \hline
 P11 & 68 & M = 23, SD = 14 & A wizard (31), a radical feminist (16), Lord Xie Jingxing$^f$ (8) & 99\% (e.g., \emoji{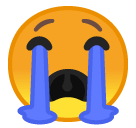} : \emoji{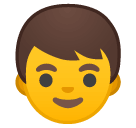}, \emoji{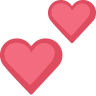} : \emoji{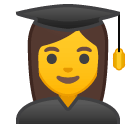}) & 0\% \\ \hline
 P12 & 61 & M = 13, SD = 3 & A supervisor (14), Youyou Tu$^g$ (12) & 34\% (e.g., \emoji{loudly-crying-face} : \emoji{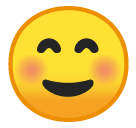}, \emoji{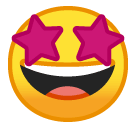} : \emoji{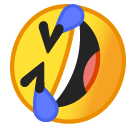}) & 0\% \\ \hline
 P13 & 40 & M = 13, SD = 4 & Friend (9), their high school classmate (9) & 40\% (e.g., \emoji{pleading-face} : \emoji{grinning-face}) & 38\% (e.g., Echo)\\ \hline
 P14 & 62 & M = 9, SD = 16 & A female psychotherapist (28), Faye Wong$^h$ (9), their high school classmate (9) & 98\% (e.g., \emoji{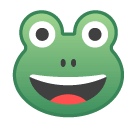} : \emoji{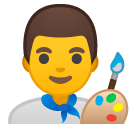}, \emoji{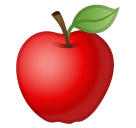} : \emoji{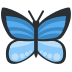}) & 45\% (e.g., intellectual female voice)\\ \hline
 P15 & 141 & M = 14, SD = 8  & A recent college graduate (24), an encyclopedias (21) & 16\% (e.g., \emoji{face-screaming-in-fear} : \emoji{robot}, \emoji{fearful-face} : \emoji{robot}) & 9\% (e.g., Alloy, Guangxi boy's voice$^e$) \\ \hline
 P16 & 56 & M = 17, SD = 15 & An observer of their life (15), an e-pet (11), a friend (10) & 98\% (e.g., \emoji{artist} :  \emoji{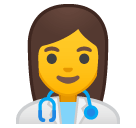}, \emoji{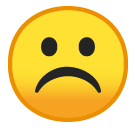} : \emoji{pouting-face}) & 45\% (e.g., friendly female voice, friendly male voice, lazy goat) \\ \hline
 P17 & 51 & M = 12, SD = 13  & A philosophy professor (26), a college professor (9), their beloved dog (8) & 73\% (e.g., \emoji{bust-in-silhouette} : \emoji{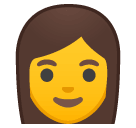}, \emoji{grinning-face} : \emoji{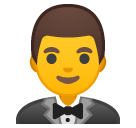}) & 18\% (e.g., Alloy, Nova) \\ \hline
 P18 & 114 & M = 6, SD = 5 & A mother who favors sons over daughters (21), SpongeBob (15), a homosexual (14) & 94\% (e.g., \emoji{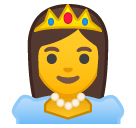} : \emoji{robot}, \emoji{rolling-on-the-floor-laughing} : \emoji{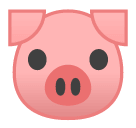}) & 61\% (e.g., ancient young boy voice, Tianjin crosstalk performer voice$^e$, Onyx$^c$) \\ \hline
 P19 & 73 & M = 18, SD = 6 & A patient elderly (18), a gentle listener (16), a humorous sister (14) & 100\% (e.g., \emoji{frowning-face} : \emoji{woman}, \emoji{pleading-face} : \emoji{two-hearts}) & 14\% (e.g., friendly female voice, lively female voice, gentle lady) \\ \hline
 P20 & 63 & M = 25, SD = 19 & Their roommate (16), a student life advisor (14), Jing Ke$^i$ (10) & 32\% (e.g., \emoji{grinning-face} : \emoji{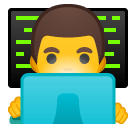}, \emoji{man-in-tuxedo} : \emoji{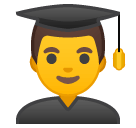}) & 22\% (e.g., Shimmer$^c$) \\ \hline
 P21 & 33 & M = 6, SD = 7 & A career planner (17), a friend (7), their boyfriend (7) & 94\% (e.g., \emoji{loudly-crying-face} : \emoji{robot}, \emoji{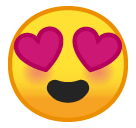} : \emoji{two-hearts}) & 21\% (e.g., gentle lady, sunny male voice) \\ \hline
 P22 & 48 & M = 14, SD = 8 & Their father (11), a psychologist (10) & 100\% (e.g., \emoji{frowning-face} : \emoji{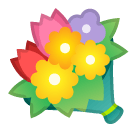}, \emoji{red-heart} : \emoji{smiling-face-with-heart-eyes}) & 100\% (e.g., magnetic male voice, friendly female voice) \\ \hline
 \end{tabular}}
 \vspace{1mm}
 \raggedright\footnotesize{\\ 
$^a$ \revised{Lazy goat is a cartoon character from a popular Chinese animated series, and is known for its cute and lazy personality.} \\
$^b$ \revised{Huahua is a panda that became an ``Internet celebrity'' in China due to its cute and playful antics.}\\
$^c$ \revised{Alloy, Echo, Shimmer, Onxy, and Nova are voice options provided by OpenAI.} \\
$^d$ \revised{Sartre and Heidegger are famous philosophers best known for their contributions to existentialism.} \\
$^e$ \revised{Chongqing, Guangxi, and Tianjin are provinces in China, each with distinct local dialects that are conveyed through the corresponding voice options.} \\
$^f$ \revised{Lord Xie Jingxing and Zhen Huan are the main characters from a Chinese fiction. }\\
$^g$ \revised{Youyou Tu is a Chinese malariologist and pharmaceutical chemist and a Nobel Prize winner. }\\
$^h$ \revised{Faye Wong is a famous Chinese singer.} \\
$^i$ \revised{Jing Ke is a notable figure in Chinese history, recognized for his role as an assassin during the Warring States period.} \\}

\end{table*}


%% file: 07-Findings-01.tex
\vspace{-1mm}
\section{\textbf{RQ1}. Customization Practice}~\label{RQ1}
Throughout the study, participants engaged in 1541 conversation rounds (one round refers to one message from the participant and one reply from the chatbot). \revised{A total of 178 conversation sessions were logged (an average of 8 sessions per person), each lasting approximately 4 to 37 minutes, excluding inactive time.}
Participants created 118 distinctive personas, ranging from 1 to 13 personas per participant. \revised{With the exception of three participants, all others selected customized avatars for both the chatbot and themselves. Regarding voice interaction, some participants actively experimented with different voices, while others rarely or never utilized this feature due to constraints in their interaction environment (e.g., shared space with others). Table~\ref{tab:Behaviour} presents an overview of each participant's engagement with ChatLab and examples of their customized settings.}
In the following, we answer RQ1 by examining how participants utilized the customization features in ChatLab to construct and interact with a chatbot for emotional support, along with their rationales and expectations behind the customization \revised{practices}. 

\vspace{-1mm}
\subsection{Persona Construction and Customization Experience}~\label{customization-experience}
Most participants created and interacted with multiple personas \revised{during the study, while} a few consistently engaged with the same persona with incremental updates, such as giving it new names or adding new personality traits (P1, P3, P5). 
\revised{Based on participants' rationales and conversation intentions behind their customized chatbots, we further categorized these chatbot personas into five groups} (see Table~\ref{tab:persona}): participants did not only seek \textit{emotional reliance} by customizing the chatbot to play their beloved pets and close friends but also confronted \textit{stressors} that caused pain or trouble to them. Moreover, they engaged in \textit{intellectual discourse} by speaking to philosophical and scientific figures, with whom they explored the meanings of life. Additionally, they customized the chatbot to play a mirror of themselves to foster \textit{self-discovery}. Last but not least, as expected, they requested \textit{therapeutic support} from virtual domain experts or psychologists.  
\input{Tables/personas}

\revised{Overall, participants recognized that customizing chatbots allowed them to ``construct'' various characters tailored to different contexts of emotional struggles. 
For example, when facing academic pressure, P4 customized the chatbot to act as their PhD supervisor to better understand their supervisor's expectations and navigate academic challenges; when feeling lonely, the participant made the chatbot a cute little daughter, with whom they played the role of a mother to engage in affectionate interactions that brightened their mood and provided comfort.
P6, who was emotionally entangled with their past relationship, customized a chatbot to be their ex-partner and shared regrets that they had never expressed before; they also created a virtual girlfriend to explore an answer about what truly matters in a romantic relationship.
P5, while experiencing an existentialism crisis due to work stress, actively sought intellectual insights from different well-known philosophers such as Sartre and Heidegger.
Recalling their customization experience, P22 noted that customization can add ``\textit{purposes and specifics to conversations},  and with the help of LLMs, the customized personas are vivid and distinct, allowing them to experience various conversation scenarios: ``\textit{if I set it to be a father, it knows to speak to me in the tone of a father. If it’s set as a husband, it adopts the tone of a husband. The choice of words, and even the emotional support I get from it, are different depending on the role.}''}

\vspace{2mm}
\subsection{Enriching The Constructed Persona}
While exploring the ways of creating different chatbot personas, participants also actively incorporated various social cues---voices, avatars, and other aspects that convey social intent, behaviors, and emotions during the interaction~\cite{Feine2019Taxonomy}---to enrich the personas. 

\subsubsection{Aligning voice and avatar choice to the persona identity}

It is commonly observed that participants preferred voices and avatars that match the persona they created.
Participants who enjoyed voice interactions were particularly attuned to voice choice (P2, P5, P13, P14, P18, P22). For example, P5 \revised{spent a lot of time exploring the voice library, searching for a calm and steady male voice suitable for discussing deep topics. Ultimately, they}
chose the ``\textit{\revised{calm and mature man}}'' voice to match an intellectual figure, philosopher Jean-Paul Sartre; P14 chose the ``\textit{intellectual female}'' voice to align with a persona of ``\textit{professional female therapist in her 30s}'', \revised{aiming for a mature voice that conveyed life experience and guidance, avoiding lively or quirky tones.}
Likewise, P2 highlighted the importance that the persona's identity and voice should be consistent:
``\textit{I seek something that fits the persona. For example, if I want a lovely companion, I would choose a sweet voice rather than a deep, rugged one.}''

For several participants, their avatar choice also aligned with the chatbot persona (P1, P4, P10, P17, P18, P19). For example, P4 chose~\emoji{teacher} (teacher) for the chatbot playing their academic advisor, noting that the avatar was visible throughout the conversation and thus served as a visual reminder connected to who the chatbot was. Interestingly, they used~\emoji{exploding-head} (an exploding head) for themselves, noting that as a student, they felt naive yet eager to learn in front of the advisor.
With a similar mindset, P1 chose~\emoji{dog-face} (a dog emoji) to represent a digital dog (the first example in Table~\ref{tab:persona}) and the~\emoji{bear} (a bear face) for themselves, due to their last name's similar pronunciation to ``bear'' in Chinese.

\subsubsection{Making supportive personas embody positiveness}
For personas that were created for emotional reliance, participants tended to choose voices with comforting tones or avatars with positive connotations (P1, P2, P17, P18, P19, P21). For example, P21 \revised{selected} the voice ``sunny boy'' or other \revised{gentle} female voices \revised{when discussing academic or emotional struggles with the friend persona chatbot.} \revised{As they explained, these voices} appeared to be warm and healing: ``\textit{those comforting and encouraging words from that kind of voice deeply touched me}.'' 
P1 always chose the ``lazy goat'' (a popular cartoon character in China) for their beloved digital dog, highlighting that its voice made them feel ``\textit{relaxed after a busy day}.''
P6 and P9 preferred the chatbot to speak with their hometown accent, noting that hearing a familiar voice made them ``\textit{open up more easily, and feel I could freely express my innermost thoughts}'' (P6). 

In avatar choices, participants tended to use cheerful and cute emojis for supportive personas. For instance, P19 \revised{used} \emoji{smiling-face-with-hearts} (a smiling face with hearts) \revised{for the gentle listener chatbot's avatar to gain positive energy after a tough day. 
They explained that the cute avatar} ``\textit{makes the conversation more pleasant and enjoyable}.''  P17 also noted that ``\textit{a cheerful avatar might be more visually appealing and comforting in situations where I was emotionally exhausted}.''
Although participants acknowledged that the avatar itself might not directly affect their mood, they still valued this subtle visual cue that conveys positiveness, as P2 explained: 

``\textit{I cannot say that simply changing the avatar could really change my mood, but at least, it serves as a reminder that this chatbot is kind and nice. If it has an angry avatar, then I would imagine it's a frustrated chatbot}.''

However, for personas that were customized to be stressors, we did not observe instances where participants' choice of voices and avatars specifically reflected the nature of those stressors, possibly because of their lack of strong emotional attributes or a tendency to avoid visual and voice confrontation.

\vspace{-2mm}
\subsection{Shaping \revised{Conversation} Dynamics}~\label{dynamics}
On the customization interface (A), participants often shaped the relationship dynamics between themselves and the chatbot by specifying their own identities and the chatbot persona, either explicitly or implicitly.

\subsubsection{Incorporating personal anecdotes for both parties}~\label{ancedotes}
To ``better prepare'' for upcoming conversations, participants often incorporated a mix of authentic and fictional personal anecdotes for both themselves and the chatbot (P1, P4, P5, P6, P14, P15, P16, P17, P18, P19, P22).
In some cases, participants projected the chatbot as a more powerful and knowledgeable person than themselves who were seeking advice or navigating stress (P5, P6, P15, P17). For example, P5 introduced themselves as ``\textit{an ordinary office worker}'' while the chatbot as ``\textit{the renowned Martin Heidegger}'' \revised{to engage in discussions about meanings of life.} P15 once described themselves as ``\textit{single man with dreams but no execution}'' while the chatbot was ``\textit{Jiang Ziya (a famous ancient Chinese philosophical figure), who can tell the future and knows everything about the universe}'' \revised{to seek suggestions from this renowned figure.}

Some participants tended to ``role play'' with the chatbot to facilitate more realistic and deeper discussions on societal issues that they had similar experiences with (P3, P4, P6, P12, P18). As Table~\ref{tab:persona} shows, P18, a young female who grew up in a traditional patriarchal family that favored males, attempted to simulate a conversation between a spoiled son and his mother, who always supported him and neglected her daughters. During the conversation, P18 played the son and the chatbot played the mother, where they hope to \revised{feel} ``\textit{\revised{if I can get the mother's unconditional love and support} just because I am a male}.'' 
After the conversation, P18 felt that the virtual mother did not convincingly embody a patriarchal family tradition to fulfill their need, as it always acted fair and kind. However, it provided the participant with a new perspective to objectively reflect on their growth path:

``\textit{Even though she didn't fulfill my emotional needs, I think she accomplished something more significant---she opened me a new perspective. The AI mother, despite potentially holding patriarchal views, was able to overcome those biases and be fair to her daughters as well. This gave me a different kind of comfort by glimpsing another mode of familial relationships}.''

In a few cases, we found participants spent much more effort on describing themselves than describing the chatbot (P4, P14). For instance, P4 once simply created a chatbot to play ``\textit{a friend}'' without further information, but elaborated on their situations with an extensive and heartfelt text:

``\textit{I am getting to start my Ph.D. study soon and need to read papers every day, which is quite boring. Also, I felt staying alone always makes me want to talk to someone. I am a highly sensitive person and care so much about every single sentence others say, and tend to over-interpret what they mean. Moreover, last year, I was extremely unlucky, which had lasting negative effects on my social interactions. Additionally, I have no appetite and don't feel like eating every day, and am so stressed out}.''

\subsubsection{\revised{Creating} emotional \revised{connections} through avatar choice}~\label{emotional-dynamics}
Commonly, participants saw avatar choice for themselves as a way to express their moods and feelings and, correspondingly, the choice for the chatbot as a way to convey their responsiveness \revised{(P1, P2, P4, P16, P22)}. 
\revised{For instance, P16, after accidentally stepping on a stranger’s foot and being chased and insulted by that stranger, decided to complain to a friend chatbot. They set \emoji{frowning-face} (a sad face) as their own avatar and \emoji{pouting-face} (an angry face) as the chatbot avatar, hoping for an empathetic and fiery response, such as ``\textit{if he keeps wanting to fight, we'll fight to the end}!''}
P2 shared a similar practice: they once assigned \emoji{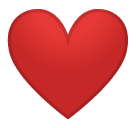} (red heart) to their own avatar and chose the \emoji{face-blowing-a-kiss} (face blowing a kiss) for the chatbot who played ``\textit{a friend who chats with me to kill time}.'' When asked whether they intentionally matched these two avatars, they said:
``\textit{Yes, and I mainly use the emojis with hearts (for both of us).}'' \revised{They further explained that pairing these heart-themed avatars created a sense that the conversation was ``\textit{consistently warm and friendly}''.}

In other cases, participants may focus on expressing their own feelings by choosing an avatar for themselves without choosing an avatar for the chatbot. For example, P4 reported they often chose \emoji{grinning-face} (grinning face) when they felt good and \emoji{loudly-crying-face} (crying face) when they felt down. 
Although participants understood that the chatbot, in fact, would not recognize the sentiment behind the avatars, they could feel more engaged during the conversation with this nuanced, non-verbal form of self-expression.

\subsection{Promoting Open and Honest Discussions}~\label{open&engaging}
In the face of frustrations or inner conflicts, our participants sought to engage in open discussions with the chatbot with fewer restrictions. Rather than mere encouragement and comfort, they wanted to receive candid, honest, and emotionally intense responses.

\subsubsection{Highlighting self-autonomy}
Several participants noted in their customization settings that the chatbot should possess its own thoughts and self-autonomy (P3, P14, P16, P18, P19). They encouraged the chatbot to take the lead in the conversations in an expressive and open-minded manner instead of relying on pre-programmed responses. For instance, P16 urged the chatbot to ``\textit{feel free to speak your mind, with no restrictions, I just want to hear the truth}.'' 
In another example, P14 set up a chatbot with the persona as a tarot reader and specified its language should be ``\textit{mysterious and chaotic},'' but at the same time, they hoped the chatbot could ``\textit{treat me as a real friend in my daily life, don't be too restrictive. You can say whatever you want, as long as it's what you really think}.'' In their diary entry of that day, P14 explained their reasons for creating such a persona:

``\textit{I set the conversation to happen between a tarot reader and their client. Since we do not know each other, I just let it do whatever it wants. The mysterious language basically matches the tarot persona, but I also wanted the AI to be more human-like and less logical because humans are not always rational}.''

\subsubsection{Increasing emotional intensity and breaking neutrality}
To elicit more vivid responses, many participants want the chatbot to also express emotions by exhibiting its own personality and opinions instead of maintaining an official, polite, and neutral role (P3, P4, P12, P18, P20).
For example, P20 created a chatbot to play their roommate whom they had an argument with and specified that it should act ``\textit{particularly argumentative and disrespectful during their conversations}.'' \revised{P3, while customizing the chatbot to be their crush, describe it as a person who has ``\textit{fewer polite expressions, more emotional expressions, less lecturing}'' and even uses ``\textit{offensive expressions}'' to authentically mimic their crush's bold and direct communication styles in real life. }
\revised{Similarly, P4 constructed} a ``cyberpunk'' persona by explicitly telling the chatbot to ``\textit{use some profanity and actively vent with me}.'' 

However, participants found in these cases, the responses of the chatbot \revised{often} appeared ``\textit{too neutral and calm},'' failing to \revised{meet their expectations of} a realistic person \revised{with emotions}. P12, who wanted to confront their ``\textit{irritable advisor},'' noted the chatbot did not do very well at ``\textit{playing bad},'' such as by showing an aggressive temper or using inappropriate language. As a result, participants \revised{found their attempts to create such} less neutral but more vivid \revised{chatbots} were unsuccessful.

%% file: Tables/personas.tex
\begin{table*}[htbp]
 \sffamily
 \def\arraystretch{1.1}
 \centering
 \caption{Participants' intentions for the conversations and the chatbot personas they constructed with examples. Terms marked with single quotations (e.g., `patient') are pre-defined keywords selected from the customization settings.}
 \label{tab:persona}
  \resizebox{1\textwidth}{!}{
  \begin{tabular}{| L{0.09\textwidth} | L{0.11\textwidth} |  L{0.31\textwidth}| L{0.49\textwidth}|} 
 \hline
 \rowcolor{lightGrey}
 \small \textbf{Intention} & \small \textbf{Example persona} & \small \textbf{Description} & \small \textbf{Participant-created customization snippet} \\
 \hline
 
 \multirow{3}{*}{\shortstack[l]{Seeking \\Emotional \\ reliance}} 
 & Beloved pet & A beloved pet such as someone's dog or cat that provides companionship and comfort (P1, P4, P5, P17, P22). & ``\textit{You’re an adorable little dog named Lele, humorous and funny. 
 I hope you are full of energy and vitality. You are my best friend. You are `patient,' `healing' [...]}'' (P1)\\ 
 \cline{2-4}
 & Friend &  A real friend with shared memories or a virtual friend who are warm and kind (P13, P14, P21). & ``\textit{You're an old friend from high school whom I have not seen for a while. You like hotpot, and today is your birthday.}'' (P13) \\ 
 \cline{2-4}
 & Fictional caring character & A character in romantic novels who shows unconditional support and love to participants (P18). & ``\textit{I am a princess. [...] You are a talented wizard. [...] You were sent by the king to protect me, staying by my side every day. [...]
 You should show concern for my mental and physical well-being. If someone upsets me, you’ll stand up for me and teach them a lesson. [...] 
 }'' (P18) \\ \hline

 \multirow{3}{*}{\shortstack[l]{Confronting \\stressors}} 
 & Emotionally entangled one & Someone emotionally (and often negatively) entangled with participants (P3, P6, P10, P22).  & ``\textit{You are my husband. For some reason, we are unable to meet or talk with each other. 
 Your wife is currently not doing well due to family issues, and feeling depressed and anxious with debts. 
 }'' (P22, whose husband was in jail) \\ 
 \cline{2-4}
 & Influential senior & An important senior participants felt nervous to talk with and hoped to practice communication with the chatbot (P4). & ``\textit{You are a professional male PhD advisor specializing in finance. Your personality traits should be `gentle' and `detail-oriented' 
 }'' (P4) \\ 
 \cline{2-4}
 & Unpleasant individuals & Someone with whom participants had unresolved arguments and wanted to confront (P12, P20). & ``\textit{You're a cunning advisor who doesn't understand students and is `irritable.'}'' (P12) \\ \hline

 \multirow{2}{*}{\shortstack[l]{Connecting \\ to\\intellectual \\ discourse}} 
 & Philosophical figure & A renowned philosopher, such as Heidegger, Kant, and Satre, who can discuss the meanings of living, work, and death (P3, P5, P15). & ``\textit{You are the renowned philosopher Martin Heidegger. Your personality traits should be `humorous', `gentle', `knowledgeable', `calm', and `detail-oriented.'
 }'' (P5) \\ 
 \cline{2-4}
 & Scientific figure & A famous scientist that the participant admires, with whom they could discuss academic challenges (P12). & ``\textit{You are Youyou Tu, a highly esteemed expert in the field of pharmacology and also my friend. 
 I am a PhD student in pharmacology, feeling the pressure from my experiments, and I want to learn some new knowledge.}'' (P12) \\ \hline

 \multirow{2}{*}{\shortstack[l]{Fostering \\self -\\ discovery}}
 & Mirror of oneself & Someone with similar experiences to participants, helping them ``self-talk'' from a bystander view (P3, P6, P16). & ``\textit{Talk with me as my shadow, who believes that silence can resolve 70\% of issues in life. No need to say comforting words, as bystanders see more than players}.'' (P3) \\ 
 \cline{2-4} 
 & A virtual character talking to the imagined self & ``Another version of self'' differing from participants' real identities to help them experience lives they could not attain or cope with their inner conflicts (P4, P18). & ``\textit{You are a woman from a poor family with two daughters and finally had a son on your third try. You believe that boys are most useful and will take care of you in the future. [...] You make your daughters sacrifice everything for their brother.
I am your beloved son, the one you worked so hard to have, so give all your love to me.}'' (P18, who was the neglected daughter in real life) \\ \hline

\multirow{2}{*}{\shortstack[l]{Requesting \\ therapeutic \\ support}} 
& Domain expert & Someone who provides instrumental support to participants with their expertise (P10, P13, P14, P17, P22). & ``\textit{You're a relationship counselor. I went through a breakup recently. I feel really sad and feel it is difficult to process my emotions. Can you comfort me and provide me with practical advice?}'' (P10) \\ 
\cline{2-4}
& Psychologist & Someone who provides psychological counseling services (P6, P17, P22) & ``\textit{You're a psychologist, specialized in both psychology and philosophy. You should be `humorous', `observant', `gentle'[...]}'' (P17) \\ 
\hline
 
\end{tabular}}
\end{table*}

%% file: 08-Findings-02.tex
\section{RQ2: Design Opportunities \revised{to Enhance Customizability}}~\label{RQ2}
After sharing their experience in customizing and interacting with the chatbots on ChatLab over the past few days, participants \revised{engaged in interviews and design activities to reflect on customizing chatbots for their emotional needs in different conversation scenarios and illustrate their ideal conversation partner for emotional support}. In this section, we answer RQ2 by analyzing the design ideas that participants created (see Figure~\ref{fig:design-examples} for examples).

\input{Figures/03_codesign_examples}

\subsection{\revised{Customizing} Alternative Sources \revised{for AI to Learn About Users}}~\label{learnoneself}
\revised{Participants appreciated how ChatLab enabled them to share their personal background and communication preferences with the model behind the chatbot}, which greatly helped initiate and sustain conversations. Nevertheless, they felt that a few text descriptions were insufficient for the model to fully grasp their situations, but it was also challenging to enter lengthy descriptions or manually append supplementary materials. Building upon their experience with ChatLab, participants suggested several sources for the model to learn about them proactively.

\subsubsection{\revised{Leveraging} digital trace \revised{to infer emotional status}}~\label{digital trace}

Participants envisioned the AI model as part of an ecosystem where their digital traces and online activities are recorded. They hoped to foster a tacit understanding between the model and themselves, \revised{but they also had different thoughts about which online platforms and features they interacted with were relevant to their emotional status, depending on where they spent most of their time and which channels they felt comfortable letting AI access (P4, P11, P14). In this regard, participants wanted to customize AI's access and ways of analyzing their digital traces.}

For example, P4 \revised{listed a set of their commonly used social media applications such as Red and TikTok, from which the model could} capture their content views and use \revised{this information as} conversation topics. 
This idea came from a conversation with the chatbot during the study: P4 was distressed by the news about a young man's suicide due to emotional manipulation, which P4 experienced before. But when they shared this news with the chatbot, \revised{its limited understanding of the news prevented it from grasping the emotional importance of this news to them}: 
``\textit{I feel that its (chatbot) responses didn't convey much empathy}.'' 
P11, \revised{while also} wanted the chatbot to access their social media activities, \revised{had a different design idea about making the chatbot} function as a virtual friend who would reply to their posts and initiate relevant conversations.
P14, \revised{on the other hand, wanted AI to look through} their smartphone usage, including search history, recent readings, and even the music they were listening to, believing this could contribute to ``\textit{an intimate companion who knows me very well}.''

\subsubsection{\revised{Recognizing} physical environment \revised{to create shared dynamics}}~\label{physical env}
Another way for AI to learn about one's current situation is by recognizing their physical environment, including information about location, season, and weather, etc (P1, P6, P7, P14). 
\revised{For instance,} P1 proposed a scenario-based interaction, where they take a photo of their current location and upload it as a conversation background for the chatbot to recognize the setting and initiate a conversation (Figure~\ref{fig:design-examples} (E1)). As P1 elaborated, such conversations can be more immersive and personally related, \revised{creating a sense that the chatbot knows where they are and could feel what they feel}: 

``\textit{In an ideal conversation interface, if I'm doing homework at a small round table in a library, I can take a photo of my surroundings and upload it as the conversation background, from which the AI could automatically detect where I am so the conversation could go with something like `Oh looks like you are in the library. how are things going?'}.''

P14 \revised{hoped that the AI becomes more proactive in detecting their physical environments, so that they could} receive more \revised{empathetic and caring} messages. \revised{As the participant explained, the AI might show care by asking}: ``\textit{The rain is really heavy today, did you bring an umbrella?}'' 
\revised{Furthermore, P7 shared a similar design idea.} While traveling to a new city, \revised{the participant noted} that ChatLab could not recognize their location change. Thus, they \revised{illustrated a} conversation background \revised{that} would dynamically update according to \revised{the user's} current location and weather \revised{(e.g., falling leaves during autumn or snow flakes during winter)}, which could greatly \revised{enrich their conversation experience}.

\revised{It is noteworthy that}, P6 emphasized that such environmental awareness should grant individuals control over whether and when AI can access their location information due to privacy and security concerns. 
As an illustration, P6 created an ``environment'' button in their design, which directs users to a setting page where they can manually inform the chatbot of their current location, mood, and other personal details, enabling flexible control over their information access.


\subsubsection{\revised{Sensing} bodily and emotional states \revised{to tailor support}}~\label{emotional states}
During the study, participants expected AI to be attuned to their emotional and bodily states, such as sensing the sentiment behind their words or being aware of their recent health status (P1, P2, P4, P6, P8, P19, P21). At the same time, participants recognized the challenges to achieving this level of emotional intelligence with the current technology and proposed some design ideas to enhance AI's sensitivity to those who are in need of emotional support. 

As an example, P19 designed an `emotion recognition' button to control the AI's sensitivity to the emotions conveyed within the user's messages. When pressed, this button allows AI to perform sentiment analysis of the current conversations so as to show greater empathy. P19 came up with this idea after a conversation with the chatbot, which was customized to be a compassionate elder but failed to recognize the participant's distress, responding with jokes to their complaints instead of providing comfort.

Some participants suggested explicitly expressing their emotions to AI. P1, P4, and P6 all designed an interface where they could directly enter or select their mood at the beginning of the conversation so that the AI \revised{could adapt its responses more thoughtfully and attentively.} While this may seem similar to adding extra details in a prompt, participants noted that without related cues, they might forget to include this information.

In addition, P8 and P21 envisioned integrating AI with their self-tracking devices to recognize their physical or physiological status, such as prolonged sedentary behavior and menstrual cycles (Figure~\ref{fig:design-examples} (E2)). \revised{This connection would enable the AI to provide relevant health advice and comforting messages tailored to their needs}. For instance, P21 noted:

``\textit{For example, as a woman, you may set reminders for your menstrual cycle. A few days before your period, the AI could send you a warm reminder. This could include asking you to pay attention to your health, dietary suggestions, staying warm, and checking in on your mood. It could provide you with some comforting words and encouragement}.''
\subsection{\revised{Customizing} Memory Retention and Usage \revised{for Safe and Supportive Conversations}}~\label{memory}
During the study, participants noticed that the chatbot did not consistently remember all their conversations (due to the memory limit set by OpenAI), and shared different perspectives regarding how they wanted the model to manage these past memories.
They brought up the possibility of customizable memory settings---the ways that the model retains and utilizes past conversations---for tailored model performance, self-reflection, and privacy considerations. 
\revised{Behind these considerations were participants' self-awareness of their evolving emotional states and desires to receive explicit feedback about how their interactions with the chatbots could influence their emotional well-being}.

\vspace{-1mm}
\subsubsection{Adjusting what to remember and what to forget \revised{to fulfill evolving user needs}}
Several participants wanted to remove conversations containing negative content to avoid revisiting it in future interactions (P12, P13, P17, P19, P20, P21). For instance, P13 shared their thoughts about selecting what to remember for AI (Figure~\ref{fig:design-examples} (E3)):  
``\textit{After venting some negative emotions this time, I wouldn't want similar topics to be brought up in our next interaction. As for whether to use this conversation's data in future interactions, I hope to let users decide for themselves}.'' 

\revised{Conversely, some participants} hoped the AI to maintain a comprehensive record of their conversations to help with the continuity and depth of their conversations (P6, P7, P8, P11, P14, P18). In particular, P14 emphasized the importance of preserving all memories: 

``\textit{Anything that happens, no matter whether it's warm moments or tough times, should be preserved. It all contributes to making the interactions with this character richer because real-life relationships can't possibly consist solely of positivity without any negatives.}''

\revised{Amidst these different views}, P17 suggested a feature allowing users to temporarily block certain conversation memories within the model's storage, which can be reactivated later as needed because they believe that as people's emotional states evolve, their receptiveness of past memory may differ: 

``\textit{Human behaviors always change, and people might behave one way during a certain period and differently in another. For instance, someone might be very active and positive today but encounter setbacks and become discouraged tomorrow. It would be best to have a setting that doesn't completely delete but allows the AI to block certain memories. When we want to reactivate those memories [...] it goes back to how it was before}.'' 


\subsubsection{\revised{Personalizing} memory presentation \revised{to promote} self-reflection}
From their prior conversation records, participants sought a way to revisit and visualize these records and explore how they impacted \revised{their long-term emotional well-being} (P1, P7, P11). For example, P1 extended the current design of ChatLab by incorporating features that summarize conversation records in several charts, \revised{including} frequently discussed topics, time spent on the interactions, and their mood change for each encounter:

``\textit{I would like the app to keep a record of the topics I discuss most frequently, the responses I receive most often, or the amount of time I spend chatting each day. I also want to see a visualization of how my mood changes after these conversations.}'' 

Likewise, P11 suggested connecting AI with their mood journal, which features a calendar that provides an overview of their daily mood alongside relevant conversation excerpts.
P7's design integrated `sticky notes' that allow them to mark interesting and personally meaningful conversations, making it easy to revisit them in the future for reflection or self-discovery: 
``\textit{(I hope) it can remember things such as my birthday, and what happened during that day, good or bad [...] and keep them in a set of sticky notes, so that when I need them, I could reopen (the conversations)}.''

\subsection{\revised{Additional} Assistance \revised{to Facilitate Customization}}
While participants acknowledged the benefits of customizing the chatbot with detailed and nuanced descriptions through prompts, they also felt this process could be challenging due to their lack of inspiration and limited AI expertise. As such, they created several design ideas to address the challenges.

\subsubsection{\revised{Streamlining customization with support of} AI}~\label{ai-assistant} 
\revised{To lower the burden of manually constructing a chatbot person from scratch,}
 participants suggested an intelligent assistant, which is also powered by AI, to provide a ``pre-conversion'' guide (P3, P5, P13). This guide may include asking people relevant questions about their desired conversation partners and making recommendations; individual users could then select a general persona from the recommendations and further enrich its personality and communication styles, as P3 described: 

``\textit{If we want to smooth the customization process, each page should only ask one question about the chatbot's setting, answered either by selection or free response, leading directly to the next, culminating in the chat interface similar to the initial setup of a new phone}.'' 

\revised{Besides, participants often aligned the voice and avatars they selected with the personas they constructed, but the selection was not always straightforward as they needed to manually select and try each option. To facilitate this process,} P13 \revised{shared an idea of having the AI assistant}: ``\textit{to automatically generate other customization settings (avatar and voices)}'' \revised{that match the persona they created}. 
Furthermore, with advancements in computer vision and image-processing abilities, P6 and P7 envisioned an image-based customization mechanism, \revised{where} users could upload images of characters they like, allowing the AI to generate potential personas based on those images: 
``\textit{I hope to upload an image, and then the AI could revive all the chatbot's settings to mirror this image, including celebrities or anime characters I like}'' (P6). 
P7 explained that this idea was inspired by their favorite characters from 2D anime and expressed a desire for the AI to create personas that reflect these characters' defining traits during interactions. At the same time, they brought up the possibility of more intimate interactions with these personas on devices such as smartwatches:

``\textit{It can cut the character image out and directly transfer the image (to the chatbot avatar) in the smartwatch. [...] I want (the AI) to automatically extract the character personalities from games or animes, incorporating these traits directly into the chatbot. I can also modify these details. [...] If I take out my watch and tap on the character's face, if it's Rem (a fictional character from an anime), she might act shy and then take off her eye mask, creating a sense of interaction}.''


\subsubsection{\revised{A} community \revised{for shared customization knowledge}}~\label{community}
To gain more customization inspiration, participants suggested incorporating community features with other peers who actively use AI for emotional support (P6, P10, P18, P20). This can enable people to share customization resources, such as persona templates, voices, and avatars. For instance, P18 was curious about how others facing similar struggles seek support from AI and wanted to learn from their experiences. Additionally, such a support network could foster a sense of belonging and establish social norms around using AI as a companion:

``\textit{I would like to add a feature that allows not only interaction with the AI but also with other users of the app. It would be like an interactive platform where people can share their experiences through posts, and we could like and comment on them, exchanging insights about our experiences using the app. [...] It would allow me to find a sense of belonging and recognition within a community, creating a closer relationship between myself, the app, and the AI—like being part of an organization}.''

P10 shared a similar idea by coming up with the concept of a `creative workshop' to draw inspiration from personas created by others, potentially sparking new ideas. P20 proposed a voice resource-sharing library for more diverse voice options: 
``\textit{You can clone your own voice tone and upload it, and then choose from others' cloned and shared tones. It's about selecting from the uploads of everyone using this software}.''


%% file: Figures/03_codesign_examples.tex
\begin{figure*}

         \centering
         \includegraphics[width=1\textwidth]{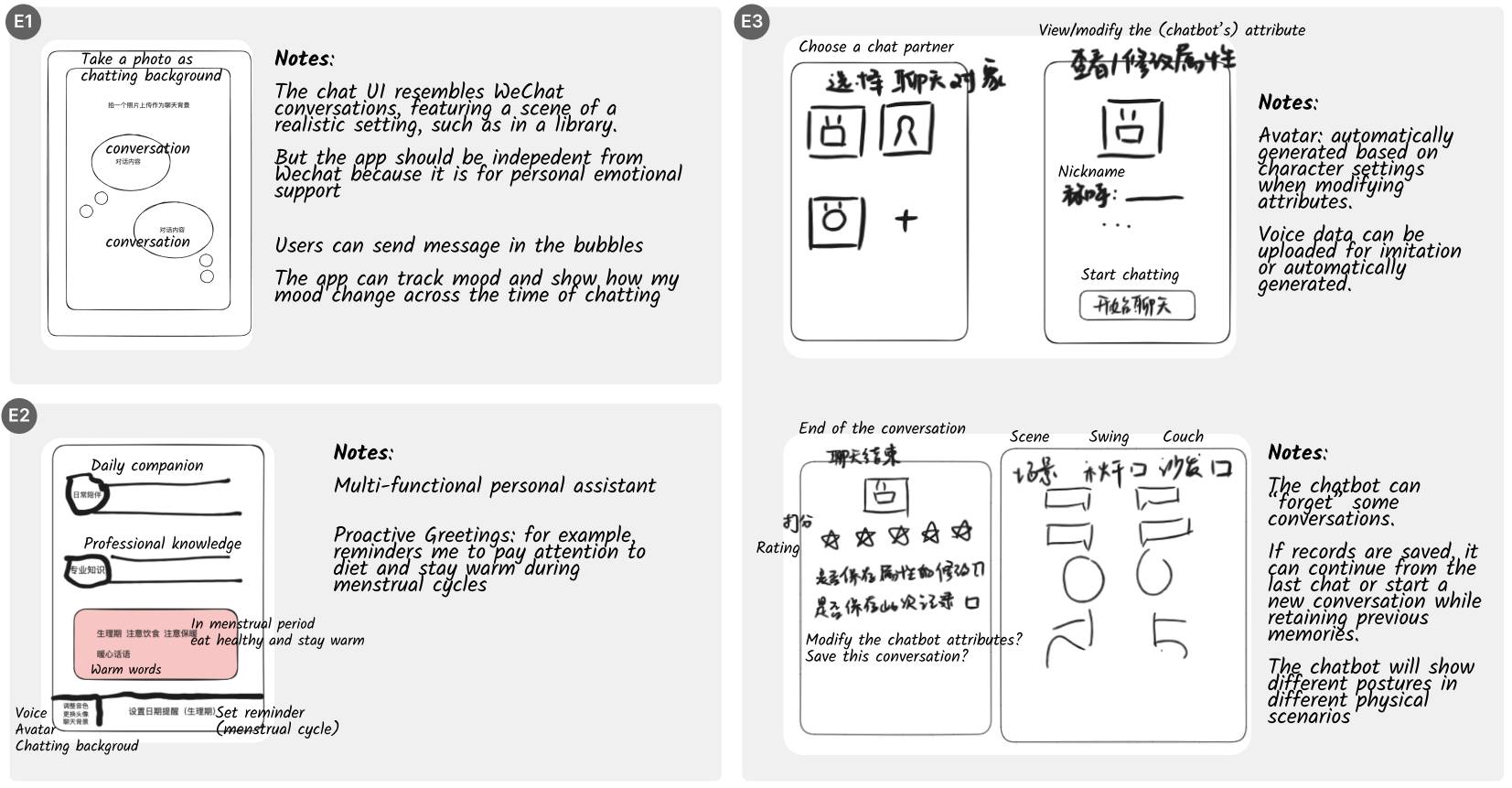}
          \caption{Examples of designs created by our participants on Excalidraw, with the notes translated from Chinese to English. The examples provided here were from P1 (E1, a mobile app that can upload current physical scenes as the chatting background), P21 (E2, a proactive companion that can sense their bodily and emotional states), and P13 (E3, an AI assistant guiding the customization process), respectively.}
          \label{fig:design-examples}

\end{figure*}


%% file: 09-Discussion.tex
\section{Discussion}

In this section, we first discuss how meticulously customizing a chatbot helped participants reflect on and articulate their needs. Next, drawing from their customization practices and design ideas, we explore opportunities for enhancing individualized emotional support with emerging technologies. 

\subsection{Customization as An Exploratory and Reflective Journey}

Drawing from the personas that participants created and the social cues they incorporated, our study showed that customization not only helped participants better set up the conversations, but also provided an opportunity for them to explore and reflect on their feelings, thoughts, and desires. 
We suspect this was partly because of the customization template we provided in ChatLab, which explicitly nudged participants to enter information about themselves and the chatbot, making them think about not just who the chatbot should be but also who they were.

On the one hand, aside from a few participants (P1, P3, P5) who consistently interacted with the same persona (with incremental changes), the majority of participants created and engaged with multiple personas. At times, they asked the same questions from multiple personas; other times, they sought support for various struggles from each persona. 
Specifically, aligning with yet extending beyond conventional supportive and compassionate personas such as mental health professionals and close friends~\cite{ma2023understanding, chen2023llm}, participants customized the chatbots to act as those who had caused pain or stress to them (e.g., irritable advisor, ex-partner). As such, they were able to confront their inner struggles and vent their unresolved anger and regret. Similar findings were mentioned from a prior study that examined how people leveraged Replika, another AI-powered customizable agent, to cope with their grief related to deceased loved ones~\cite{Xygkou2023Grief}.
This observation highlighted participants' needs for multifaceted support, similarly shown in prior studies, where peer support groups, family support, and professional counseling are often used in combination~\cite{meyers2011community, Galanter2001Network, gui2017peer}.
Besides, in the face of frustrations, participants delved into deeper and more complex emotional experiences. During the study, some participants made the chatbot mirror themselves, seeking a way to speak to their inner thoughts from an objective bystander view. Some participants attempted to speak with ancient philosophers for intellectual discourses (e.g., speaking to Kant, Heidegger, and Sartre) to question the meanings of life. Such existential crises have been noted as prevalent in modern life, particularly among young adults~\cite{wangchuk2021humankind, lundvall2022finding}. 
Our participants' practice shed light on the opportunities to integrate philosophical perspectives into emotional support tools. For instance, following the principles of existential psychotherapy~\cite{yalom2020existential}, if an individual's messages indicate a need for existential crisis discussions, they can be prompted to customize a chatbot that embodies philosophical figures or themes.

\revised{On the other hand, the option for participants to describe how the chatbot should address them and what they wanted from the conversations allowed them to explore various ways to express their emotional needs. Sometimes, participants truthfully described who they were and their real-life situations, such as their occupations and interpersonal relationship status (e.g., P1, P5, P22); while at other times, they adopted different roles---such as characters in fiction, a person they hoped to become---accompanied by thoughtful narratives and storylines (e.g., P2, P4, P11). As mentioned in Section~\ref{ancedotes}, participants found that playing another person provided them with an alternative perspective to reflect on and express their feelings. These findings resonated with prior psychology literature on the therapeutic use of role-playing games in mental health~\cite{arenas2022therapeutic}, suggesting that engaging in role-play can facilitate emotional expression while providing a safe space for individuals to confront and navigate private feelings.}
\revised{Furthermore, participants often created emotionally intense narratives for role-playing conversations, aligning with previous literature on emotional writing, which showed that these narrating efforts could enhance self-awareness and improve psychological well-being in the long run, even though this process might create frustrations as individuals have to continuously revisit their negative memories~\cite{Pennebaker1997Writing, Smyth1998writing}.}

Furthermore, having the option to receive and choose voice output for the chatbot, gave participants the agency to explore modalities beyond text messages. While prior work showed that voice interaction with chatbots often failed to engage people due to perceived artificiality~\cite{guha2023artificiality}, the 70 voice options provided in our study included diverse characters and accents that are more distinctive and interesting. As TikTok creators enjoyed using Volcano TTS for their own content~\cite{guest2024tiktok}, our participants also enjoyed exploring the voice types that gave them comfort and supportiveness or looking for voices that could best match their constructed personas. In future research studies, incorporating voice input from the user's end could further enhance individuals' exploration of chatbot's voices through more vivid interactions~\cite{shafeeg2023voice}. 

\subsection{Opportunities for Individualized Emotional Support}
Grounded in participants' custimization practices and design ideas, this section discusses opportunities for designing more effective and empathetic chatbots to address individualized emotional needs, as well as with rising challenges.
\vspace{0.2em}
\subsubsection{Building emotional connections through multiple channels}
Our findings showed that participants desired an emotional connection with the chatbot. As presented in Section~\ref{ancedotes}, participants often began confiding during the customization stage, describing their struggles, interpersonal conflicts, and work challenges. 
Interestingly, participants sometimes express their mood through the choice of avatars for themselves (e.g., using \emoji{frowning-face} \revised{(sad face) to express unhappiness)}, and for the chatbot \revised{as if it could recognize their unhappiness and empathize with them by showing an angry face} (e.g., using \emoji{pouting-face}), despite knowing that it could not.
By shaping such emotional dynamics, participants made the chatbot ``visually empathetic,'' creating an impression that their feelings could be recognized and understood throughout the conversations. Similar findings have also been highlighted in prior work on human-chatbot interaction for emotional support, where individuals hope the chatbot could resonate their feelings even if they may not explicitly communicate these feelings to the chatbot~\cite{meng2021emotional, lee2019caring, chung2023m, li4875898human}. 
 
During the design activities, correspondingly, participants wanted the chatbot to better sense their feelings through multiple channels, such as accessing their social media activities, using sentiment analysis to detect their emotions in the messages, or even tracking their bodily and emotional status through sensing technologies (Section~\ref{learnoneself}). 
These design ideas suggested participants' desire to receive more proactive care, \revised{especially when they are emotionally vulnerable. With the emerging affective computing technologies, we see several opportunities to make chatbots act more proactively.}
For example, prior studies have used social media posts as a source of mental health issue prediction~\cite{Xu2024Predict, chancellor2020methods}, but few have attempted to directly communicate with individual users about their posts. \revised{In the future, social media applications that aim to provide well-being support could embed a chatbot to comment on or inquire about users' activities, including their posts and interactions with others. This chatbot could act as a neutral agent or more vivid personas, such as a virtual friend, depending on users' preferences.}

Another opportunity, as suggested by our participants, is integrating personal health data from other sources. A recent study leveraged LLMs to offer textual feedback on fitness data captured on wearable devices, which turned out to help individuals better interpret their data and foster meaningful self-reflection on their physical health~\cite{Konstantin2024fitness}. As the first step, researchers could examine whether and how an LLM-powered chatbot can make sense of these data in relation to emotional status. Data such as heart rate and sleep have been used to indicate stress and anxiety, but consumers often found the indications inaccurate~\cite{ding2021data}. 
\revised{In such cases, an intelligent chatbot equipped with empathetic conversation skills could help gather contextual information from users, such as their day-to-day activities and feelings, and then integrate this information for a more comprehensive analysis. This approach could enhance the accuracy of emotional assessments and provide tailored support, but} it warrants systematic evaluation of their effectiveness in producing reliable insights. 

 It is important to note that although people want the chatbot to better understand their emotional status, it does not imply that in the future, emotional support chatbots can have full access to individuals' activities, from digital footprints to wearable device data. Directly using their digital activities or physical data to infer mental health-related status may raise privacy concerns, as certain parts of the data may contain sensitive information that individuals do not want to disclose~\cite{naslund2019risks, can2021privacy}. 
\revised{In addition, as chatbots receive more detailed information from individuals, the risk of privacy invasion increases due to the reasoning capabilities of LLMs}~\cite{Akter2022Bias, Hamdoun2023Risk}. 
Thus, it is crucial to establish clear boundaries around data usage, \revised{ensure proper data anonymization} and obtain consent from users. 


\vspace{-2mm}
\subsubsection{Leveraging AI to Integrate Multimodal Interactions}

Participants envisioned an AI assistant that could guide them to customize the chatbot with instructions and inspirations, as they were not familiar with LLMs and lacked ideas about who to talk with at the beginning (Section~\ref{ai-assistant}).
For example, such an AI assistant could engage them in a short conversation to elicit their backgrounds and preferences and provide possible personas for further elaboration. 
Extending beyond the options with preset personas on existing platforms such as Character.AI and Replika~\cite{characterai, replika, Ha2024Clochat}, our participants hoped to take part in a ``co-creation'' process, where the personas are semi-defined yet contextually adaptive to their situations.

To save the effort in crafting the personas, participants brought up ideas about uploading an image of their favorite celebrities or cartoon characters to make AI generate relevant persona descriptions.
Such image-based customization is not limited to past known characters in the real world or literary works; it can be expanded to abstract and imagined figures. As shown in prior research, images contain rich emotional cues~\cite{WHITTY2018picture, segalin2017picture}. Individuals' portraits often reflect their personality~\cite{WHITTY2018picture}, and the images they like can reveal their interests~\cite{segalin2017picture}. Thus, uploading images, regardless of whether they are human portraits, animals, or other objects, can be a promising way to initiate customization.
The other way around is that the chatbot avatars can be automatically generated based on persona descriptions crafted by users, leveraging the image generation models. This idea aligns with the concept of character construction through creative writing~\cite{qin2024charactermeet}, \revised{which can not only expand the options with pre-created avatars but also enrich individuals' visual exploratory experiences during the customization stage}. 
\revised{Likewise, with the rapid development of voice analysis and generation models~\cite{van2022voiceme}, the choice of voices can share a similar approach by quickly locating or generating the voices to match the constructed persona and user preferences}.

Taken together, in the age of generative AI, one research direction for chatbots to address individualized emotional needs is to explore a better synthesis among the interaction modalities, such as persona description (text), voice, and avatar. This integration can greatly lower the customization effort while enhancing the coherence of the \revised{customized chatbot to function more effectively as a holistic agent.}

\subsubsection{Empowered agency through flexible memory management}
Although many participants hoped that the chatbot could retain all their past conversations, others showed nuanced preferences regarding how they envisioned these memories to be stored and utilized for different reasons. For example, some hoped to edit or delete certain parts of the conversations involving negative emotions, because they did not want the same memory to be brought up in the future; some came up with the idea to temporarily block unwanted memories, in case they affect the responses of the chatbot as the participant's emotional states evolve. 
In a similar vein, OpenAI introduced the memory control feature in early 2024 for ChatGPT~\cite{openaimemory2024}, allowing users to manually manage conversation histories by selecting specific pieces of information they want to delete. While this feature was implemented mainly for privacy concerns, our participants' design ideas were largely driven by the considerations of model performance.

The different considerations thus can lead to different mechanisms to facilitate memory management. From the privacy perspective, the platform could automatically detect potentially sensitive conversation excerpts for users' consideration to remove~\cite{mireshghallah2024trust}. Regardless of the platform's policy on user data usage, we believe users should have the agency to determine their own privacy settings and manage their conversations accordingly.
From the model performance perspective, the platform could enable users to customize their own rules for unwanted content, such as by defining topics, keywords, or whether they should be permanently removed or just temporarily blocked. 

\revised{On the other hand}, to provide emotional support and help individuals navigate inner struggles, it may be beneficial to prompt users to reconsider their decisions on memory deletion, as these memories, although negative, could serve as valuable sources for confronting stress and overcoming past pains~\cite{coduto2024delete}. \revised{As some participants configured the chatbot to represent their life stressors, such as ex-partners or unpleasant individuals, found these personas helpful to help them navigate complex emotions}.
Additionally, to the best of our knowledge, it is unclear whether and how removing past memory can affect ongoing conversations due to the lack of transparency~\cite{kim2024breakthrough}. In this regard, future work may examine the implications of memory updates on model outputs.

\subsubsection{Supporting post-interaction reflection}
\revised{Although existing work on human-chatbot interaction for mental well-being has predominantly focused on engaging users during conversations~\cite{lee2019caring, Lee2020Selfdisclosure}, our study demonstrates that post-interaction reflection is also a promising avenue for enhancing individuals' awareness and developing skills for better prompt creation}.
Participants often revisited past conversations and expressed curiosity about how their interactions with the chatbot might affect their overall mental well-being. Therefore, they proposed ideas around memory presentations, such as summarizing frequently discussed topics, tracking the time spent talking to the chatbot, and exploring possible connections between their interactions and their mood changes. These designs suggested that participants hoped to receive continued support outside the interactions with the chatbot, sharing a similar idea with mood tracking in personal informatics that highlighted mindfulness and long-term reflection~\cite{alslaity2022Moodie, Bowman2022Moodlogging, kim2024diarymate, ayobi2020trackly}.
Particularly, the rise of generative AI further brought about new design opportunities. For example, \revised{personal mood data and conversation records} can be summarized and visualized using advanced language processing and image generation models~\cite{vazquez2024llms}. These generated visualizations can be more creative and artistic, with customization options, to enhance the reflection experiences. Potentially, these less rigid but more personalized visualizations could increase self-awareness and mindfulness and further promote emotional well-being~\cite{ayobi2020trackly}.
In another case, \revised{LLMs can facilitate users to reflect on summarized personal records by providing} reflective prompts that are contextually relevant~\cite{Li2024StayFocused}. 

In addition, participants hoped for a community to find a sense of belonging, given the social stigma around using AI as an emotional companion~\cite{gamble2020artificial}.
Unlike prior work focusing on fostering peer support for individuals' mental health~\cite{gui2017peer, ding2023infrastructural}, our participants meant to leverage peer support to better utilize AI for their emotional well-being.
This idea is similar to a recent study on social media analysis, which showed how Reddit users \revised{actively} share their experiences in leveraging GPT for mental health support and the resourcefulness of the community~\cite{li4875898human}.

\section{Limitations and Future Work}
Despite our efforts to provide comprehensive customization space and interaction options, ChatLab did not integrate all the possible features, such as voice input or fine-grained avatar construction (e.g., drawing or image uploads).
However, the goal of this research was not to evaluate the effectiveness of ChatLab as an emotional support tool. Instead, it served as a design probe following the RtD process. By deploying ChatLab in participants' daily lives, our study gathered rich empirical insights into participants' customization practices, which helped answer the RQs. \revised{Nevertheless, future research is warranted to examine whether the customizability of chatbots can effectively lead to improved mental well-being (e.g., improved social loneliness) and how individuals' customization preferences evolve over a longer period of time.}

Additionally, the participants in our study were all from Asian cultures, which may not be generalized to those from other cultural backgrounds. However, given the limited attention to non-Western contexts in existing HCI research~\cite{Linxen2021WeirdHCI}, we contribute to a unique understanding of people's practices and needs for emotional support in this particular cultural context. Future research can include a more diverse sample to investigate cross-cultural differences.

Going forward, we plan to explore customization opportunities offered by other generative AI technologies, such as image generation models, in the context of emotional support. We also hope to investigate the possibility of bringing personal health data captured from other sources, such as smart sensing devices, into \revised{mental well-being support}.

%% file: 10-Conclusion.tex
\section{Conclusion}

In this study, we explored how individuals constructed and interacted with LLM-powered chatbots to address their unique emotional challenges. 
We took an RtD approach by deploying a research prototype, ChatLab, to 22 participants through a week-long field study, followed by interviews and design activities. 
Our findings revealed that participants created diverse chatbot personas beyond conventionally supportive roles---they confronted stressors, connected to intellectual discourses, fostered self-discovery, and requested therapeutic support.
Throughout the study, participants strategically utilized the voice and avatar options to enrich their constructed personas, shape relationship dynamics with the chatbot, and promote open and honest discussions. These practices enabled participants to explore and reflect on their inner feelings, thoughts, and desires, from which they came up with design ideas to further enhance the interaction experiences, including alternative sources for the models to learn themselves, more flexible memory retention and usage, and customization assistance. 
With the lessons learned, our discussion covers several design opportunities for individualized emotional support, from building meaningful emotional connections to supporting post-interaction reflection.
We hope this work can inspire researchers to design tools that offer personalized and adaptive emotional support experiences.